\documentclass[floatfix,secnumarabic,amssymb,nobibnotes,nofootinbib,aps,pra,showpacs,aip]{revtex4-2}
\usepackage{epsfig}
\usepackage{amssymb}
\usepackage[colorlinks=true, citecolor=red, linkcolor=blue, urlcolor=cyan]{hyperref}

\usepackage{times}
\usepackage{amssymb}
\usepackage{amsmath}
\usepackage{mathrsfs}
\usepackage{graphicx}
\usepackage{cleveref}
\usepackage{dcolumn}
\usepackage{color}
\usepackage{bm}
\usepackage{graphics}
\usepackage{graphicx}
\usepackage{braket}
\usepackage{soul}
\usepackage{subfigure}
\usepackage{epstopdf}

\graphicspath{{./figure/}}
\DeclareGraphicsExtensions{.eps}

\begin{document}

\title{Time- and frequency-domain two-particle correlations of a  driven dissipative Bose-Hubbard model}
 \author{Kingshuk Adhikary$^1$, Anushree Dey$^1$, Arpita Pal$^2$, Subhanka Mal$^1$ and Bimalendu Deb$^{1}$}
 \affiliation{$^1$School of Physical Sciences, Indian Association for the Cultivation of Science, Jadavpur, Kolkata 700032, India.\\
 $^2$School of Physics and Astronomy, Rochester Institute of Technology, Rochester, NY, 14623, USA.}
\begin{abstract}
  We theoretically investigate the time- and frequency-domain  two-particle correlations of a driven dissipative Bose-Hubbard model (BHM) at and near a dissipative phase transition (DPT). 
  We compute Hanbury Brown-Twiss (HBT) type two-particle temporal correlation function $g^2(\tau)$ which, as a function of time delay $\tau$, exhibits oscillations with frequencies determined by the imaginary part of Liouvillian gap. As the gap closes near a transition point, the oscillations at that point dies down. For parameters slightly away from  the transition point, the HBT correlations show oscillations from super-bunching to anti-bunching regimes. We show that the Fourier transform of HBT correlations   into frequency domain provide information about DPT and Liouvillian dynamics.  We numerically solve the  many-body Lindblad master equation and calculate Wigner distribution of the system in steady state to ascertain DPT.
   Below certain  drive strength, the Fourier transform shows a two-peak  structure while above that strength it exhibits either a Lorenzian-like single-peak structure or a  structure with two-dips. The width of the single-peak structure is minimum at the phase transition point and the peak of this structure always lies at zero frequency. The positions of the two symmetrical peaks in case of 
two-peak structure are given by the imaginary parts of the Liouvillian gap while their half width at half maximum (HWHM) is given by the real part of the gap. The positions and the widths of the two dips are also related to  low lying eigenvalues of the Liouvillian operator. We discuss quantum statistical properties of the model in terms of the HBT correlation function and its Fourier transform.
\end{abstract}

\maketitle
\section{Introduction}
In quantum mechanics, dissipation of a system is usually treated by system-bath Liouvillian  dynamics, capturing an interplay between unitary evolution and decay processes that result from  a coupling between the system and its environment which is known as bath or reservoir.  For such system-reservoir interacting cases, dissipation is to be considered as a boundary for coherent dynamics, leading to a doorway for open quantum systems. An open quantum many-body system may be viewed as an out-of-equilibrium counterpart of an equilibrium system. The methods of theoretical exploration of open quantum systems are not as well-established as those of idealized closed quantum systems. 
Nevertheless research into dissipative or open quantum systems over the years has led to the development  of several  theoretical formalisms such as Gutzwiller \cite{PhysRevB.44.10328} and cluster mean-field \cite{PhysRevLett.97.187202,clus1,clus2}, corner-space renormalization \cite{corner1,corner2}, full configuration-interaction Monte Carlo \cite{PhysRevLett.109.230201,monte}, Keldysh formalism \cite{PhysRevLett.110.195301,Keldysh1,Keldysh2}, matrix product operator and tensor-network techniques \cite{ten1,ten2,ten3,ten4} etc.

In recent times, a number of theoretical studies on quantum phase transition (QPT) in a variety of physical platforms \cite{Zoller2,Zoller1,Casteels16,Casteels17,ciuti13,ciuti14,Biondi2} under nonequilibrium situations have been carried out. More specifically, dissipative  many-body quantum phenomena \cite{Zoller1,Zoller2} have been studied using cold atoms \cite{Baumann1,Baumann2,Esslinger,Carmichael1}, spin ensembles \cite{Keldysh2,corner2,corner3}, Josephson junctions \cite{two1,santra2}, superconducting circuits \cite{Carmichael2,Wallraff,Houck}, semiconductor \cite{Jacqmin,Rodriguez,dpt}, interacting polaritons in a Kerr nonlinear cavity \cite{polariton:2006}. Three decades ago, pioneering theoretical work on QPT between superfluid (SF) and Mott-insulator (MI) was carried out by Fisher's group \cite{Fisher}. Subsequently, experimental demonstration of QPT has been reported \cite{Greiner,Baumann1,Takahashi1} in an optical lattice loaded with an atomic Bose-Einstein condensate. For ultracold atoms in traps or optical lattices, various kinds of losses can be controllably generated with external fields or particles, leading to dissipative engineering of  driven many-body  quantum systems \cite{dissipative:engineering}. One-body particle loss \cite{santra2} is implemented by applying electron beam in a controlled manner. Two-body loss \cite{Takahashi1,Takahashi2,Syassen08,Rempe} is a fundamental property of a many-body system and related to inelastic collisions. In ultracold atom optical lattices, two-body loss has been engineered by controlled photoassociation \cite{Takahashi1,Takahashi2}. Three-body dissipation \cite{three1,three2,three3} has been realized by Feshbach resonance with controllable strength of three-body recombination.

In order to obtain a  dynamical or nonequilibrium  phase of a dissipative many-body system, an external field is required as a drive. In quantum optics, a coherent drive has enormous utility as a one or two-photon pump \cite{two1,two2}, opening up new vistas in 
light-matter interactions. A nonequilibrium system exhibits DPT \cite{immam12,Fazio17,biella18} when the Liouvillian spectral gap closes in some well-defined limit analogous to thermodynamic limit. Recent experimental observation of nonequilibrium phase transition or DPT \cite{dpt} in a driven system has given a tremendous impetus to the field.

One of the key issues in the context of DPT in a driven open quantum systems is the role of higher order quantum fluctuations as the system is driven towards the transition point. In particular, the study of two-particle correlations is important as they carry crucial information about the quantum statistical properties of the system. In a ramarkable  recent experiment, Fink {\it et al.}  \cite{dpt} have explored the decay dynamics of HBT type two-particle correlation function $g^{(2)}(\tau)$ as a possible signature of a DPT in a driven nonlinear optical system of cavity polaritons. They have observed critical slowing of the decay of $g^{(2)}(\tau)$ as the system is driven towards the phase transition point. Sciolla {\it et al.} \cite{kollath:prl:2015} have shown that the two-time two-particle correlations can be used   as a probe for complex non-stationary dynamics of dissipative many-body systems. The bunching of 
continuously pumped photon Bose-Einstein condensate in terms of HBT correlations has been experimentally demonstrated by Schmitt {\it et al.} \cite{weitz:prl:2014}. Casteels, Fazio and Ciuti \cite{Fazio17} have theoretically examined  the behavior $g^{(2)}(0)$  as a function of drive strength when a  nonlinear photon mode is driven towards a DPT. $g^{(2)}(\tau)$ has been also theoretically studied for a strongly pumped dissipative BHM of a coupled  array of nonlinear cavities \cite{ciuti14}. Syassen {\it et al.} \cite{Syassen08} have experimentally demonstrated that strong dissipation can inhibit loss and drive  a cold molecular  gas on an optical lattice into a strongly correlated system characterized by $g^{(2)}(0)$ which is much less than unity.

Here we carry out a detailed theoretical study on the  HBT correlations of a driven dissipative BHM. Depending on the system parameters, the correlations show oscillatory decay. We characterize the frequency of the oscillations by analyzing the Fourier transform of the temporal correlations into the frequency domain in terms of the Liouvillian spectral decomposition. To the best of our knowledge,  the oscillations in the decay of HBT correlations and their frequency charaterization in terms of the system parameters of a driven dissipative BHM have not been studied so far. This will be important to gain further insight into the role of two-particle correlations in DPT of the model. 

To ascertain  the occurrence of DPT in our model, we calculate  Wigner distribution \cite{Wigner} of the system and examine its features reflecting the steady-state quantum states. Our results show that for the parameters at which the system exhibits DPT, the oscillations in HBT correlations die down and the decay shows critical slowing, in consistent with the earlier results \cite{dpt}. The Fourier transform shows a single-peak spectral structure with the peak lying at zero frequency. Slightly away from the phase transition point, the oscillations revive and the spectral structure shows multiple peaks or dips depending on the system parameters. Our results show that, below certain  drive strength, the Fourier transform shows a prominent two-peak  structure.  As the drive strength exceeds that strength, 
Fourier spectrum  exhibits either a Lorenzian-like single-peak structure or a  structure with two-dips. We show that the width of the single-peak structure is minimum at the phase transition point. The positions of the two symmetrical peaks are found to be equal to the imaginary parts of the Liouvillian gap while their HWHM is given by the real part of the gap.  We discuss in some detail the quantum statistical properties of the model in terms of the HBT correlation function and its Fourier transform and highlight their characteristic features at or near the DPT.  
 
The paper is organized as follows. We describe our theoretical methods for a generic driven dissipative BHM   in Sec. \ref{sec2}. The results and their interpretations are presented in Section \ref{sec3}.  
Finally, in Sec. \ref{sec6}, we draw conclusions, highlighting the future prospects of our study.

\section{Theoretical methods}\label{sec2}

\subsection{The model and its solution} 
The Hamiltonian of a driven Bose-Hubbard model ($\hbar =1$) is $\hat H = \hat H_{ BH} + \hat H_{ drive}$ where 
\begin{equation}
 \hat H_{ BH}=-\frac{J}{z}\sum_j\left(\hat b_j^{\dagger} \hat b_{j+1}+\rm {H.c.}\right)+\frac{U}{2}\sum_j\hat b_j^{\dagger}\hat b_j^{\dagger}\hat b_j\hat b_j+ \sum_j\epsilon_0 \hat b_j^{\dagger} \hat b_{j}
 \label{eq1}
\end{equation}
is the standard Bose-Hubbard part with $\hat b_j$ and $\hat b_j^{\dagger}$ representing  the bosonic annihilation and creation operators acting on {\em j}th site. Here $J$ is the hopping coefficient between  nearest-neighbor sites, $z$ is the coordination number, $U$ is the on-site interaction parameter. The last term on the right hand side of the above equation denotes the on-site term with  $\epsilon_0$ being the on-site single-particle energy which is assumed to be same for all sites. For a system of coupled nonlinear cavities, $\epsilon_0 = \hbar \omega_c$ where $\omega_c$ is is the cavity frequency.   In case of equilibrium Bose-Hubbard physics of massive particles on a lattice, this on-site term is usually absorbed into the chemical potential. The driving  part $\hat H_{ drive}$ is given by 
\begin{equation}
 \hat H_{ drive}(t)=\sum_j\left(F \hat b_j^{\dagger}e^{-i\omega_p t}+F^* \hat b_je^{i\omega_p t}\right) 
\end{equation}
where $F$ is the one-boson driving amplitude and $\omega_p$ is the pump frequency. To eliminate the explicit time-dependency of Hamiltonian,  we may write it in a reference frame rotating at the pump frequency $\omega_p$, leading to the effective Hamiltonian 
\begin{equation}
 \hat H_{ eff}=-\frac{J}{z}\sum_j\left(\hat b_j^{\dagger} \hat b_{j+1}+\rm {H.c.} \right)+\frac{U}{2}\sum_j\hat b_j^{\dagger}\hat b_j^{\dagger}\hat b_j\hat b_j-  \hbar \sum_j\Delta\omega\hat b_j^{\dagger} \hat b_{j}+\sum_j\left(F \hat b_j^{\dagger}+F^* \hat b_j\right)
 \label{eq5}
\end{equation}
where $\Delta\omega=\omega_p-\epsilon_0/\hbar$ is the detuning between the pump and the system.

The dissipation is incorporated in the dynamics  through the Lindblad master equation
\begin{equation}
 \frac{\partial{\hat\rho}}{\partial t}=-i\left[\hat H_{ eff},\hat\rho\right]+\mathcal{D}\left[\hat\rho\right]
 \label{eq4}
\end{equation}
of the density matrix $\hat{\rho}$. Here the  dissipation of the system is described by the standard superoperator  term 
\begin{equation}
\mathcal{D}\left[\hat\rho\right]=\frac{\Gamma}{2}\sum_j\left[2\hat O_j \hat\rho \hat O_j^{\dagger}-\left\{\hat O_j^{\dagger}\hat O_j,\hat\rho\right\}\right] 
\end{equation}
 where $\Gamma$ is the damping rate and $\hat O_j$ is a quantum jump operator constructed using the combination of system operators $\hat{b}_j$ and $\hat{b}_j^{\dagger}$ depending on the nature of the dissipation process. Here $\left\{ \hat{A}, \hat{B}\right \}$ denotes an anti-commutator between the operators $\hat{A}$ and $\hat{B}$.

To solve the Liouville equation we  make an approximation  by decoupling \cite{Sheshadri,Buonsante,Cole} the hopping term 
 \begin{eqnarray}
  \hat b_j^{\dagger} \hat b_{j+1} &\approx&  \langle\hat b_j^{\dagger}\rangle\hat b_{j+1}+\hat b_j^{\dagger}\langle\hat b_{j+1}\rangle-\langle\hat b_j^{\dagger}\rangle\langle\hat b_{j+1}\rangle \nonumber \\
&=& \left(\psi^{*}\hat b_{j+1}+\psi\hat b_j^{\dagger}\right)-|\psi|^2
 \end{eqnarray}
 where $\psi$=$\langle\hat b_{j+1}\rangle$ is a bosonic coherence and site-independent. Although this approximation is not fully reliable in all physical situations as pointed out in Ref.\cite{ciuti14}, it enables one to obtain good qualitative results for the phase diagram in equilibrium BHM. In this homogeneous mean-field approximation, the hopping or tunneling term is approximated, rendering the problem effectively to a single-site dynamics. However, this approximation accounts for the on-site interaction term exactly. In the momentum space, this amounts to retaining only the zero-momentum states and neglecting all finite-momentum states. So, this is a good approximation to calculate the steady-state or dynamical properties or fluctuations around steady-state at zero temperature or zero momentum when the tunneling term is small. In the context of our model, this approximation is expected to be reasonably good as long as the tunneling matrix element $J$ is not large compared to the strength of the drive. This kind of decoupling approximation is previously used to study the dynamics of a driven dissipative photonic Bose-Hubbard model \cite{ciuti13,Carmichael2}. The Hamiltonian then takes the form $\hat{H}_{eff}=\sum_j\hat {H}_0(j)$, where
 \begin{eqnarray}
  \hat H_{0}(j) &=& \beta^*\hat b_j+\beta \hat b_j^{\dagger}-\Delta\omega \hat b_j^{\dagger}\hat b_j+\frac{U}{2}\hat b_j^{\dagger}\hat b_j^{\dagger}\hat b_j\hat b_j+\frac{J}{z}|\psi|^2
 \end{eqnarray}
  where $\beta=F-\psi J/z$ represents modified drive of the system. Under the decoupling approximation, the density matrix is product separable over the site indices. So, the density matrix for
$j$-th site $\rho(j)$ is same for all sites, and henceforth for simplicity we omit the site index $(j)$ in all the operators. Within the Born-Markov approximation we then obtain the Lindblad master in the  following form
\begin{equation}
\frac{\partial{\hat\rho}}{\partial t}=-i\left[\hat H_{0},\hat\rho \right]+\frac{\Gamma}{2}\left[2\hat b \hat\rho \hat b^{\dagger}-\left\{\hat b^{\dagger}\hat b,\hat\rho\right\}\right]
\label{eq8}
\end{equation}
We numerically solve the master equation (\ref{eq8}) at steady state ($t\rightarrow \infty$) to obtain the steady-state the density matrix $\hat\rho^{ss}$.  
We use Fock basis $|n\rangle$ and obtain a set of coupled algebraic equations. Further details of our numerical method of solution are given in appendix \ref{a1}.

For small $U$, the observable quantities of our system are found to converge when the  basis set is relatively large. In contrast, for a large value of $U$, convergence happens with a small basis set. In our numerical calculations, we ensure the independence of the size of Fock basis for all our results by choosing a sufficiently large basis set. 

Since our objective is to study second order quantum correlation and its spectral characteristics at and near a dynamical or non-equilibrium phase transition, we first semi-classically determine a transition point from mono- to bi-stable regime. In the full quantum treatment, it is well-known that there is no bi-stable regime \cite{ciuti13}, but the signature of semi-classical phase transition is manifested in the quantum treatment in a different way. Towards this end, we calculate the time evolution of the bosonic coherence $\psi$ given by 
 \begin{eqnarray}
  \frac{\partial}{\partial t}\left({\hat\rho}\hat b\right)=-i\left[\hat{H}_{0},\hat\rho \right]\hat b+\mathcal{D}\left[\hat\rho\right]\hat b \nonumber                                                                                                                  
\end{eqnarray}
taking trace on both sides, we get
\begin{eqnarray}
   \frac{\partial\psi}{\partial t}&=&-i\left[F+ \left\lbrace U|\psi|^2-\left(\frac{J}{z}+\Delta\omega+i\frac{\Gamma}{2} \right) \right \rbrace \right]
    \label{eq13}
   \end{eqnarray}
 This equation resembles to single-mode Gross-Pitaevskii (GP) equation \cite{Stringari,Carusotto,Carusotto2}. The GP equation for a dilute Bose system of photons in a single mode cavity has a similar structure.
 
 
 At steady state, the value of $\psi$ is given by solving the equation
 \begin{equation}
  \psi= \frac{F}{\frac{J}{z}+\Delta\omega-U|\psi|^2+i\frac{\Gamma}{2}}
 \end{equation}
Taking modulus on both sides, we obtain a third order polynomial equation of the mean-field mean number density $n_{mf}=|\psi|^2$ which is
\begin{equation}
 U^2n_{mf}^3-2U\left(J+\Delta\omega\right)n_{mf}^2+\left[\left(J+\Delta\omega\right)^2+\frac{\Gamma^2}{4}\right]n_{mf}-F^2=0
 \label{eq15}
\end{equation}

 \begin{figure}[h!]
\centering
  \begin{tabular}{@{}ccc@{}}
     \hspace{-.2in}
    \includegraphics[scale=0.3]{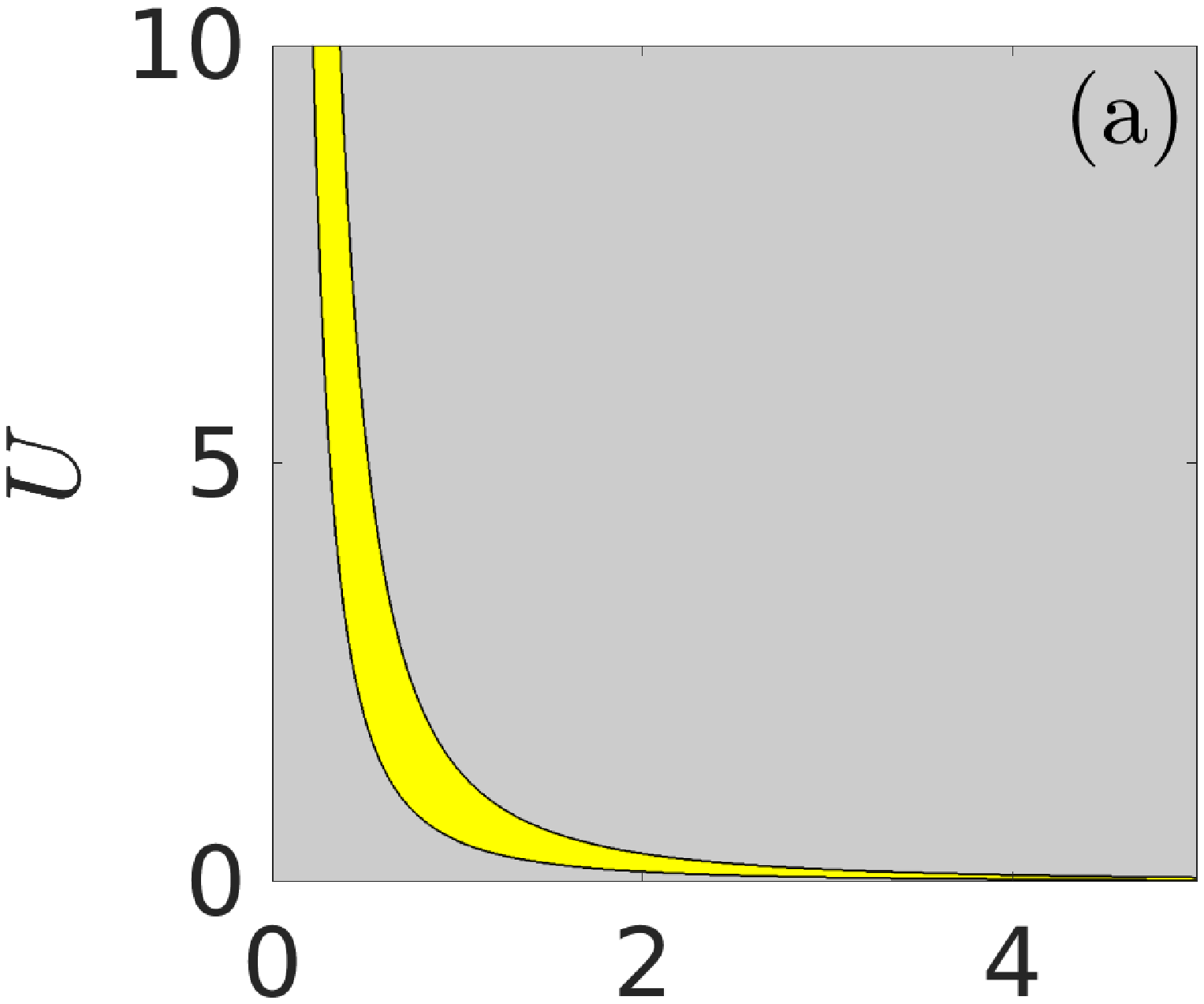} & \hspace{-.25in}
    \includegraphics[scale=0.3]{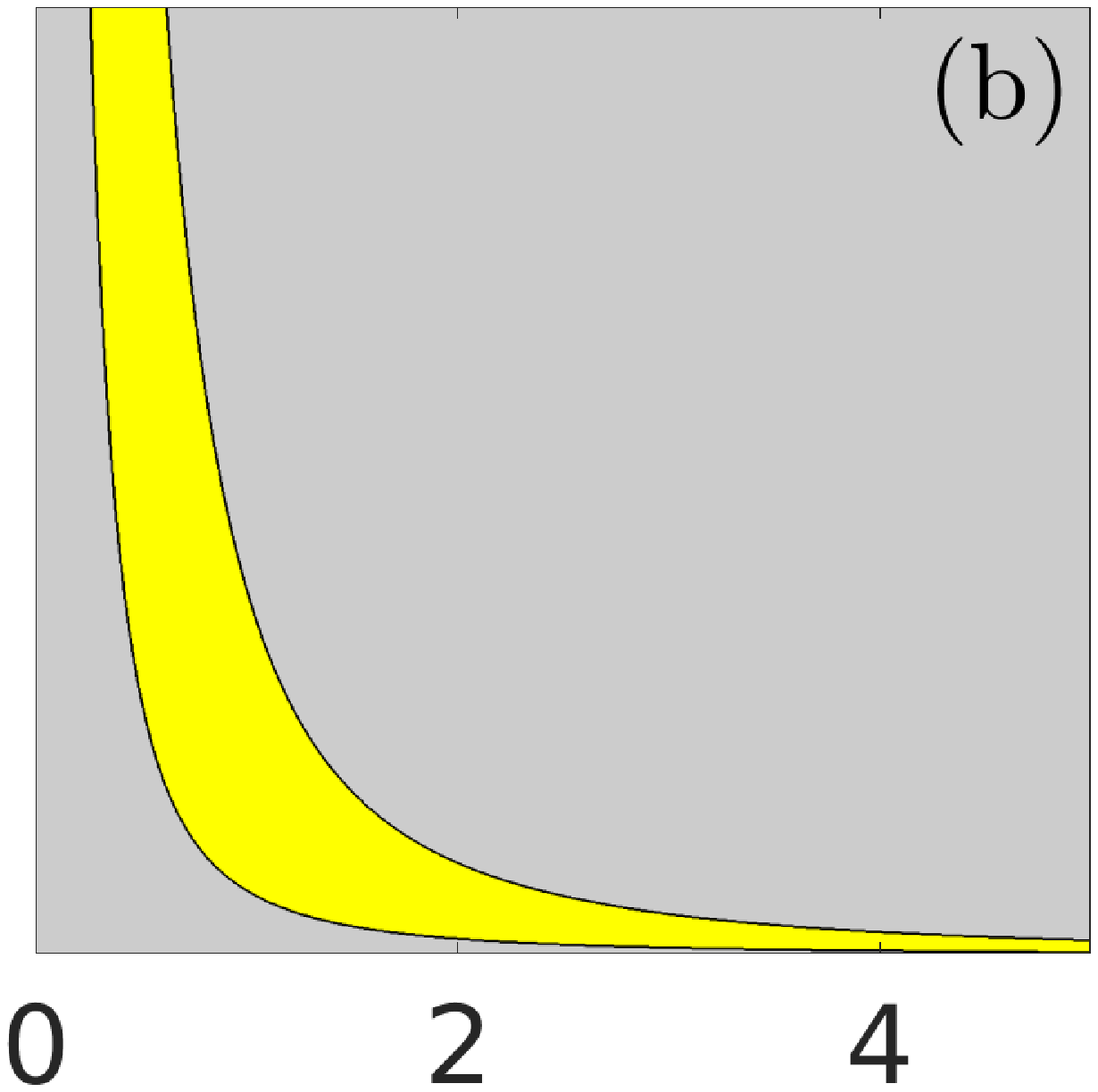} & \hspace{-.25in}
    \includegraphics[scale=0.3]{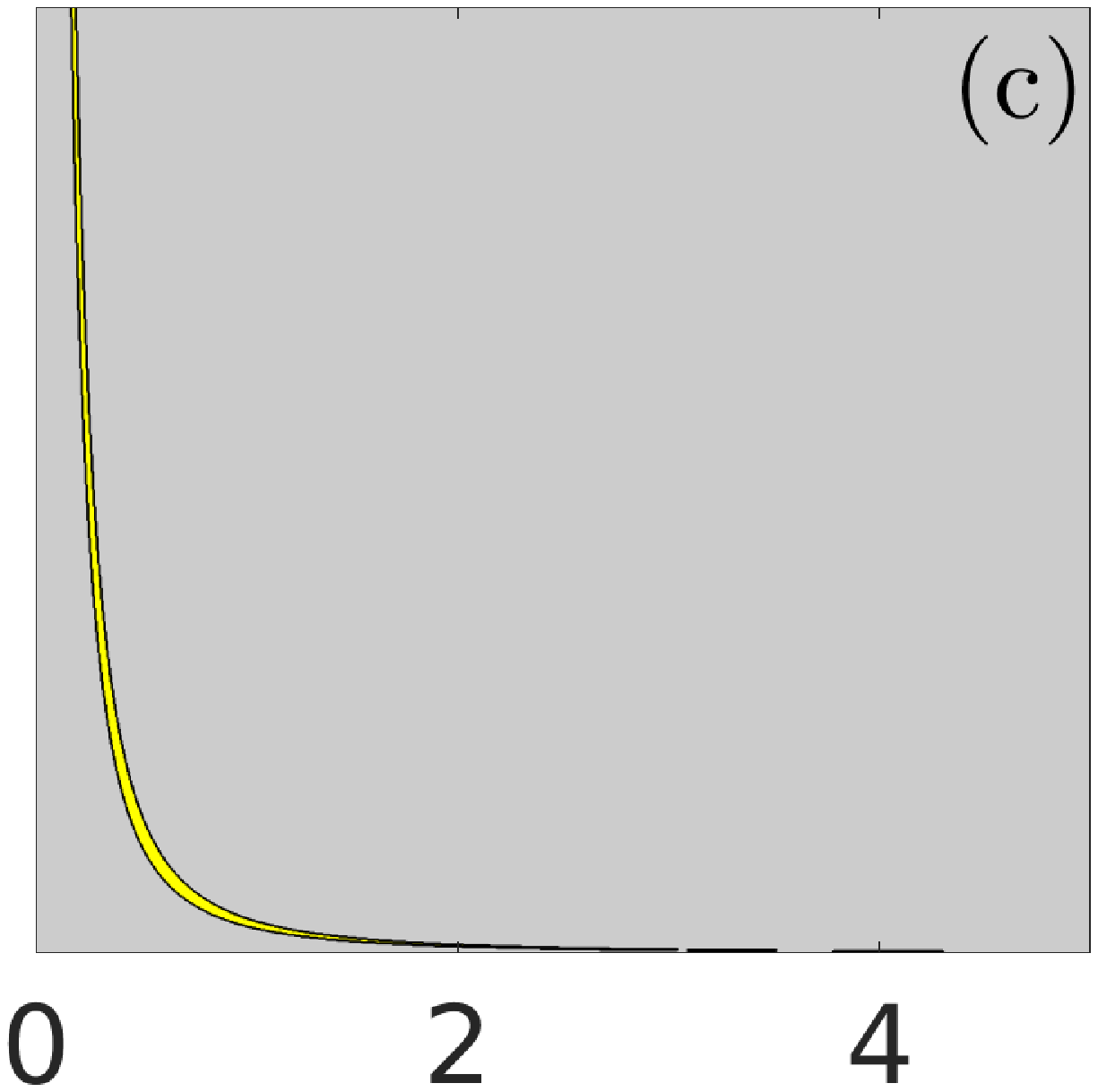} 
    \end{tabular}
\centering
    \begin{tabular}{@{}ccc@{}}
    \hspace{-.2in}
    \includegraphics[scale=0.3]{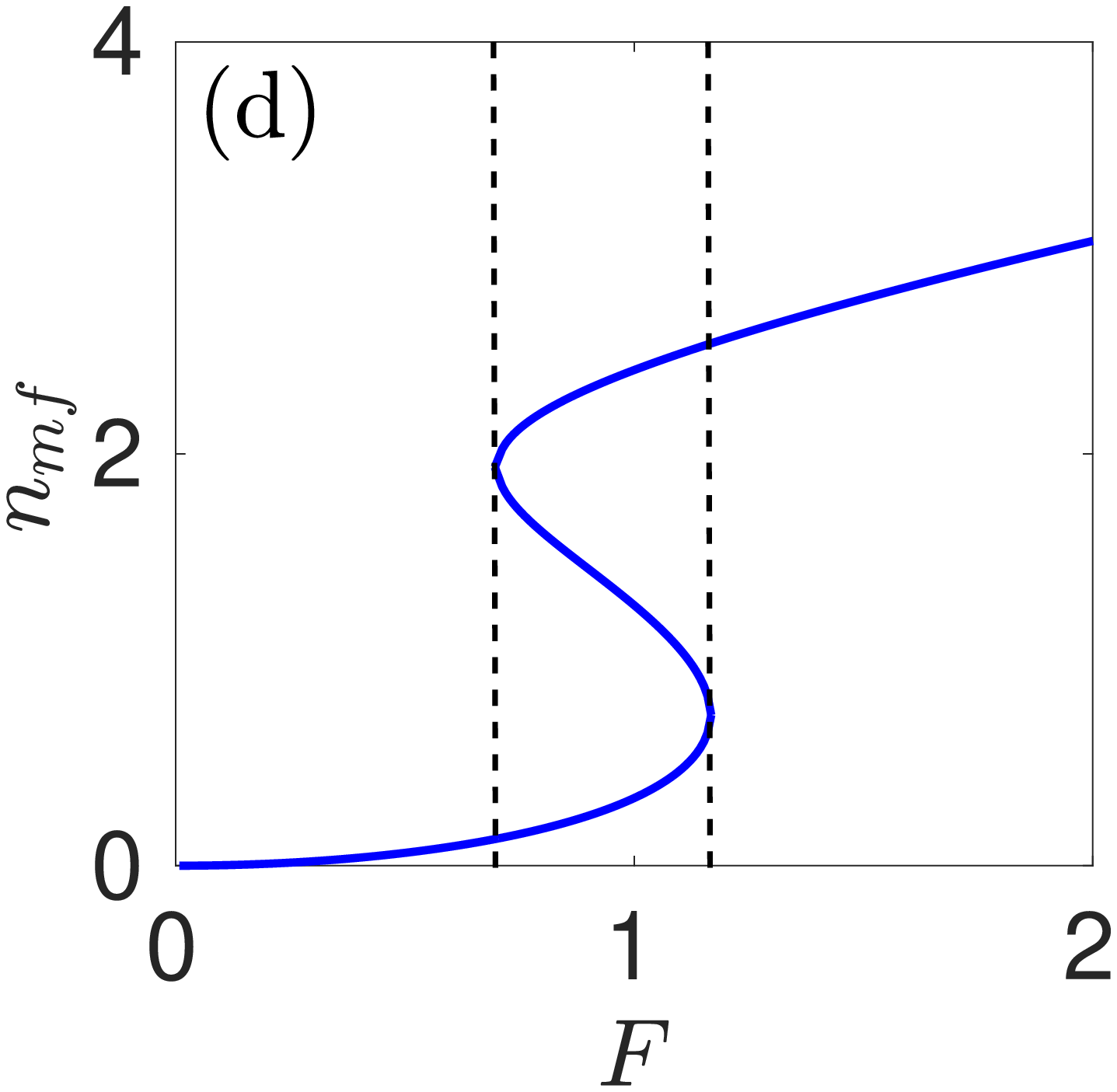} & \hspace{-.25in}
    \includegraphics[scale=0.3]{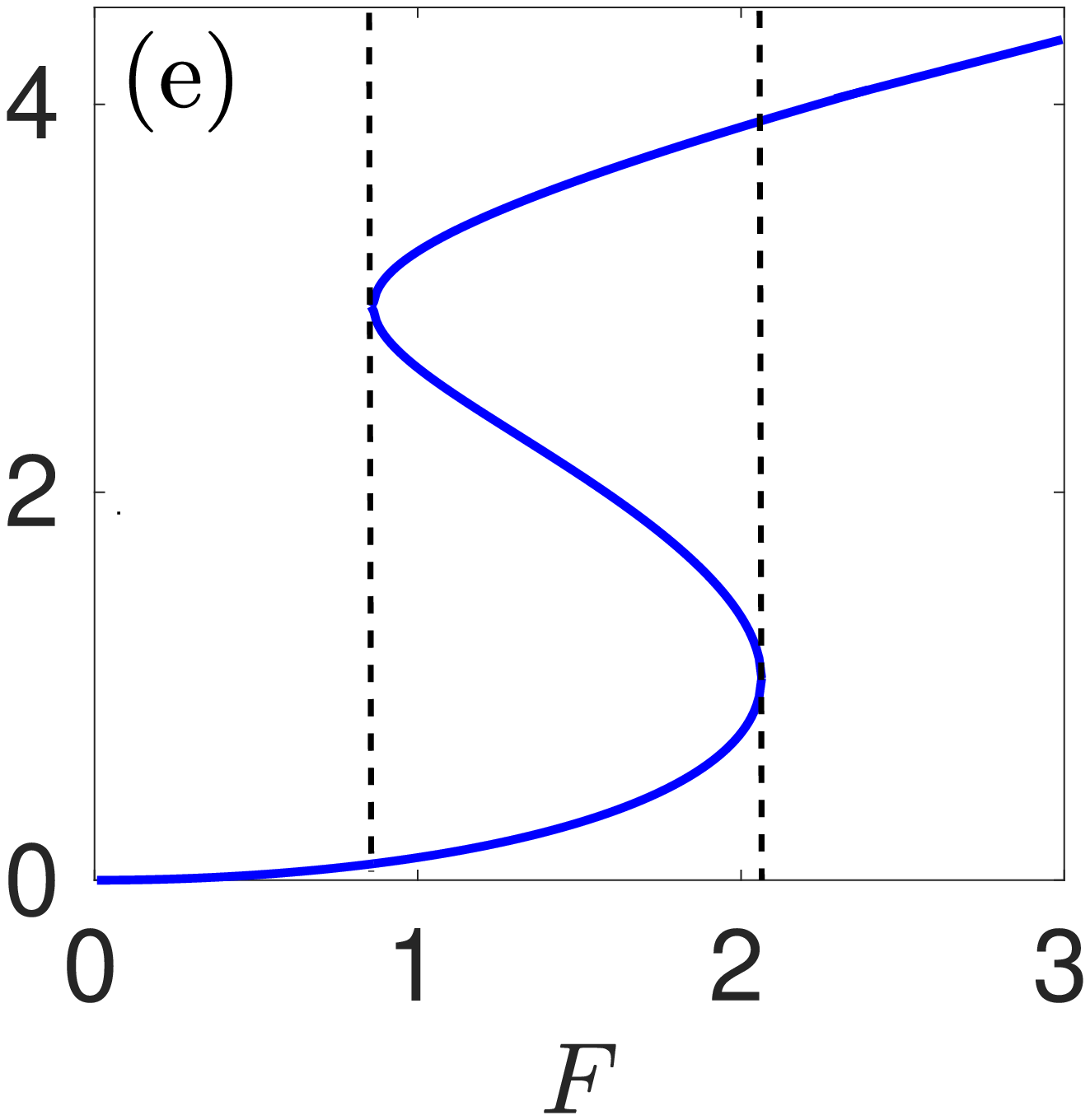} & \hspace{-.25in}
    \includegraphics[scale=0.3]{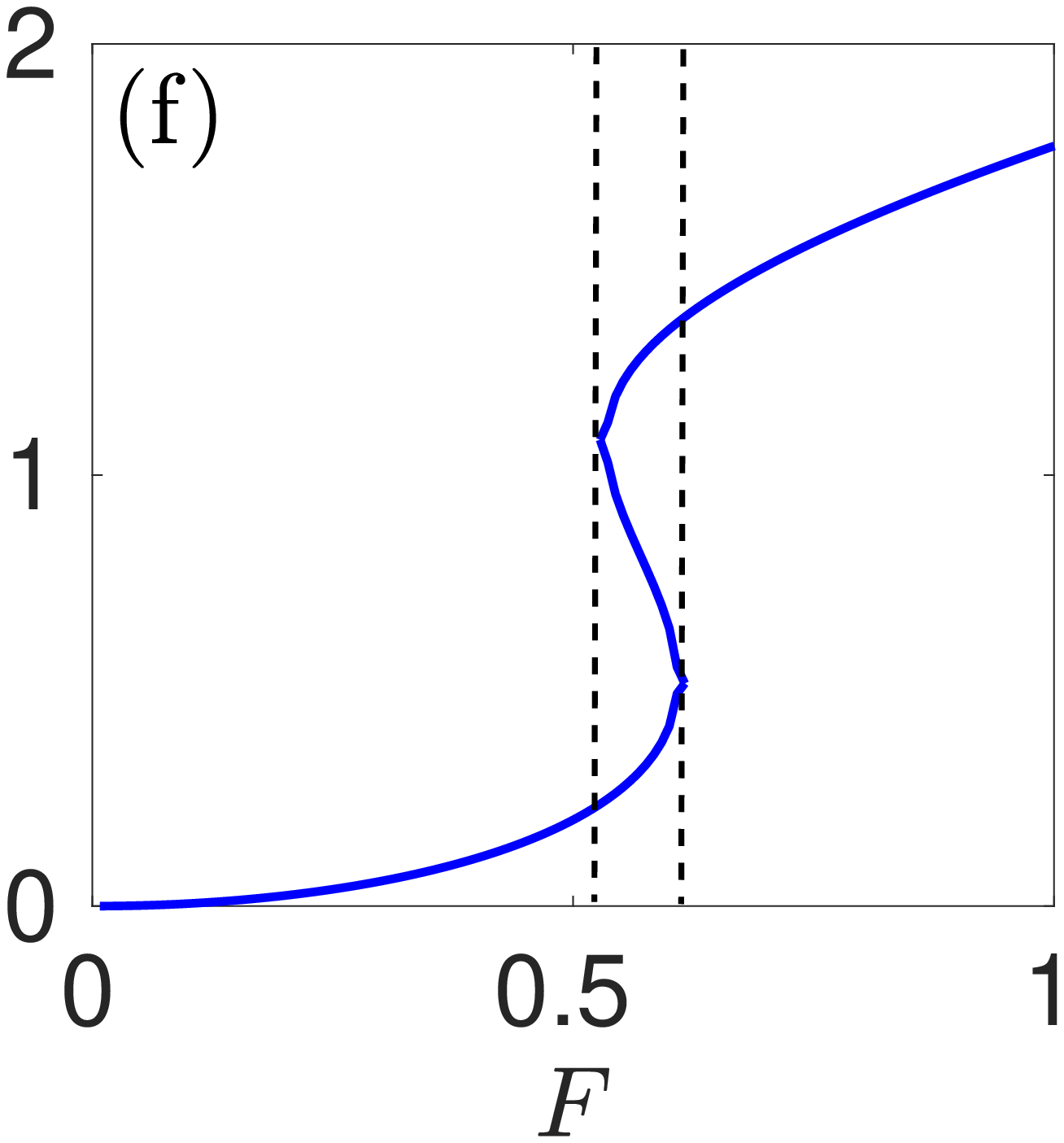}
     \end{tabular}
 \caption{ Top row represents ($U$ vs. $F$) phase space diagram of the semi-classical Eq. (\ref{eq15}) for the fixed parameters $\Delta\omega=2, J=0$ (a), $\Delta\omega=2, J=1$ (b), and $\Delta\omega=0.2, J=1$ (c). The bistable region is marked with yellow color, the gray shaded part represents the monostable region.
 Bottom row displays the variation of the mean-field density $n_{mf}$ as a function of the drive $F$ for $U=1$ with $\Delta\omega=2, J=0$ (d), $\Delta\omega=2, J=1$ (e), and $\Delta\omega=0.2, J=1$ (f).  All roots of Eq. (\ref{eq15}) are real inside the black vertical (dashed) lines.}
\label{Figure 4.}
\end{figure}

\subsection{Two-time Hanbury Brown-Twiss correlation function}
To bring forth the connection between Liouvillian spectral properties and two-particle correlations of a driven dissipative many-body system, we here briefly discuss the method of calculating HBT type two-particle correlations of the system. The evolution of the density matrix $\hat{\rho}$ governed by the Liouvillian Eq.(\ref{eq8}) can be expressed as
\begin{equation}
\frac{d \hat{\rho}}{d t} = \hat{\mathcal M} \hat{\rho}
\end{equation}
where $\hat{\mathcal M}$ is the Liouvillian super-operator. In some suitable basis, one can diagonalize $\hat{\mathcal M}$ as demonstrated by Briegel and Englart \cite{Briegel} and also by Barnett and Stenholm \cite{Barnett}.  As $\hat{\mathcal M}$ is non-Hermitian, a dual conjugate $\check{\mathcal M}$ can be constructed such that ${\rm Tr}\{\mathcal O\hat{\mathcal M}\hat\rho\}={\rm Tr}\{(\check{\mathcal M}\mathcal O)\hat\rho\}$ for an observable $\mathcal O$. $\check{\mathcal M}$ has the same eigenvalue as $\hat{\mathcal M}$. Let $u^{\mu}$ ($\mu = 1,2, \cdots$) be an eigenstate with  eigenvalue $\lambda_{\mu}$, satisfying the eigenvalue equation $\hat{\mathcal M} u^{\mu} = \lambda_{\mu} u^{\mu} $(alternatively $\check{\mathcal M}$ $v^{\mu'}$=$\lambda_{\mu'}v^{\mu'}$). A steady-state of the system corresponds to the eigenstate with zero eigenvalue. Let us denote this eigenstate by $u^{\mu=0}$. So, the steady-state density matrix   $\hat{\rho}^{ss} = \hat{\rho}(t \rightarrow \infty)  \equiv u^{0}$ is given by $\hat{\mathcal M} \hat{\rho}^{ss} = 0$. The eigenvalues with nonzero real part appear in complex conjugate pairs. The real part of a eigenvalue is non-positive and the eigenvalue whose real part has the lowest magnitude is called the Liouvillian gap.

The on-site  HBT  correlation function of a lattice is defined by 
\begin{equation}
 g^{(2)}(\tau)=\frac{\langle \hat b^{\dagger}(t)\hat b^{\dagger}(t+\tau)\hat b(t+\tau)\hat b(t) \rangle}{\langle \hat b^{\dagger}(t)\hat b(t) \rangle \langle \hat b^{\dagger}(t+\tau)\hat b(t+\tau) \rangle}
\end{equation}
The physical interpretation of $g^{(2)}(\tau)$ is that it measures the probability of detecting a particle at time $t$ and another particle after time delay $\tau$. 
For stationary processes or at steady-state of the system, the HBT function depends only on the difference $\tau$ between the two times. For  $\tau=0$, we have  equal-time second order correlation function $g^{(2)}(0)=\frac{\langle \hat b^{\dagger}\hat b^{\dagger}\hat b\hat b \rangle}{\langle \hat b^{\dagger}\hat b\rangle^2}$ which characterizes the nature of particle distribution. 
We calculate the normalized $g^{(2)}(\tau)$ at steady-state condition ($t\rightarrow\infty$) using quantum regression theorem \cite{Lax}.
Explicitly,
\begin{equation}
  g^{(2)}(\tau)=\frac{{\rm Tr}\left[\hat b^{\dagger}(0)\hat b(0)e^{\mathcal{\hat M}\tau}\left(\hat b(0)\hat \rho(t\rightarrow\infty)\hat b^{\dagger}(0)\right)\right]}{\left({\rm Tr}\left[\hat b^{\dagger}(0)\hat b(0)\hat\rho(t\rightarrow\infty)\right]\right)^2}
\end{equation}
 Here $\mathcal{\hat M}$ is the Liouvillian  matrix with infinite dimension. However, to numerically calculate the eigenvalues, we truncate the matrix upto $N^2$ such that if we increase $N$  the results remain convergent.  Here $N$ is the total number of Fock basis states.
We define a function $Q(\tau)=g^{(2)}(\tau)- g^{(2)}(\infty)$. Since ${\rm lim}_{\tau \rightarrow \pm \infty} g^{(2)}(\tau) = 1$, the on-site number fluctuation will be reduced below standard quantum limit when $Q(\tau=0)< 0 $, implying sub-Poissonian bosonic statistics. We define frequency-domain \cite{Cirac,Betzholz} two-particle correlation function by 
\begin{equation}
{\cal F}(\omega) = \Gamma \int_{-\infty}^{\infty} Q(\tau) \exp[ i \omega \tau] d\tau = \Gamma \left [\int_{-\infty}^{\infty} g^{(2)}(\tau) \exp[ i \omega \tau] d\tau - 2 \pi  g^{(2)}(\infty) \delta(\omega)\right] 
\label{ft}
\end{equation}
As derived in the Appendix \ref{a2}, we have 
\begin{equation}
 {\cal F}(\omega) =   2 \Gamma \sum_{\mu = 1}^{N^2} \left [ \frac{W_{\mu} |\lambda_{\mu r}| }{(\omega + \lambda_{\mu i})^2 + \lambda_{\mu r}^2 } \right ]
 \label{fu15}
\end{equation}
where $\lambda_{\mu i}$ and $\lambda_{\mu r}$ are the real and imaginary parts, respectively, of the eigenvalue $\lambda_{\mu}$, and $W_{\mu}$ is a weight factor as defined in the appendix.

\begin{figure}[t]
\centering
  \begin{tabular}{@{}ccc@{}}
    \hspace{-.2in}
    \includegraphics[scale=0.3]{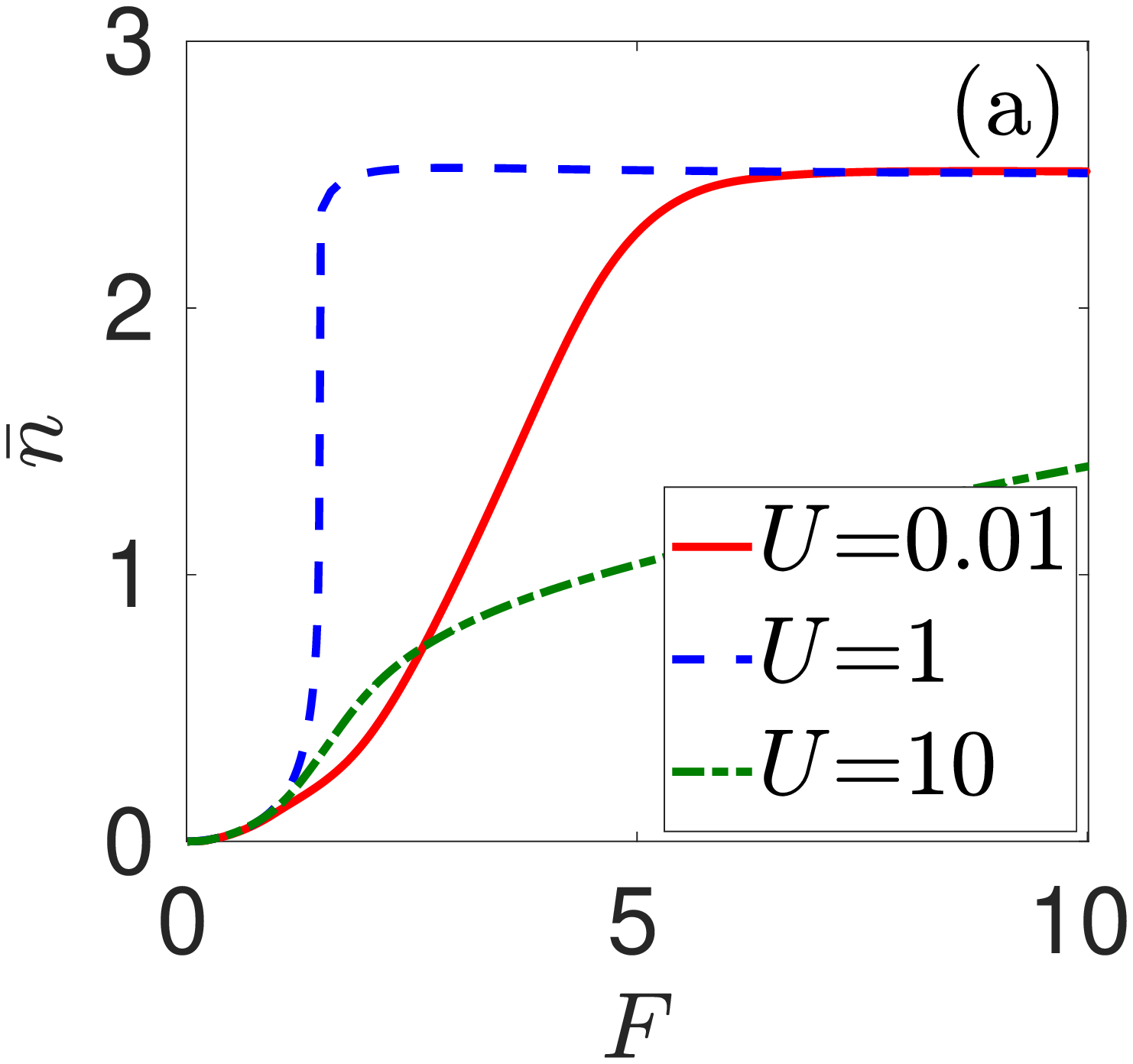} & \hspace{-.2in}
    \includegraphics[scale=0.3]{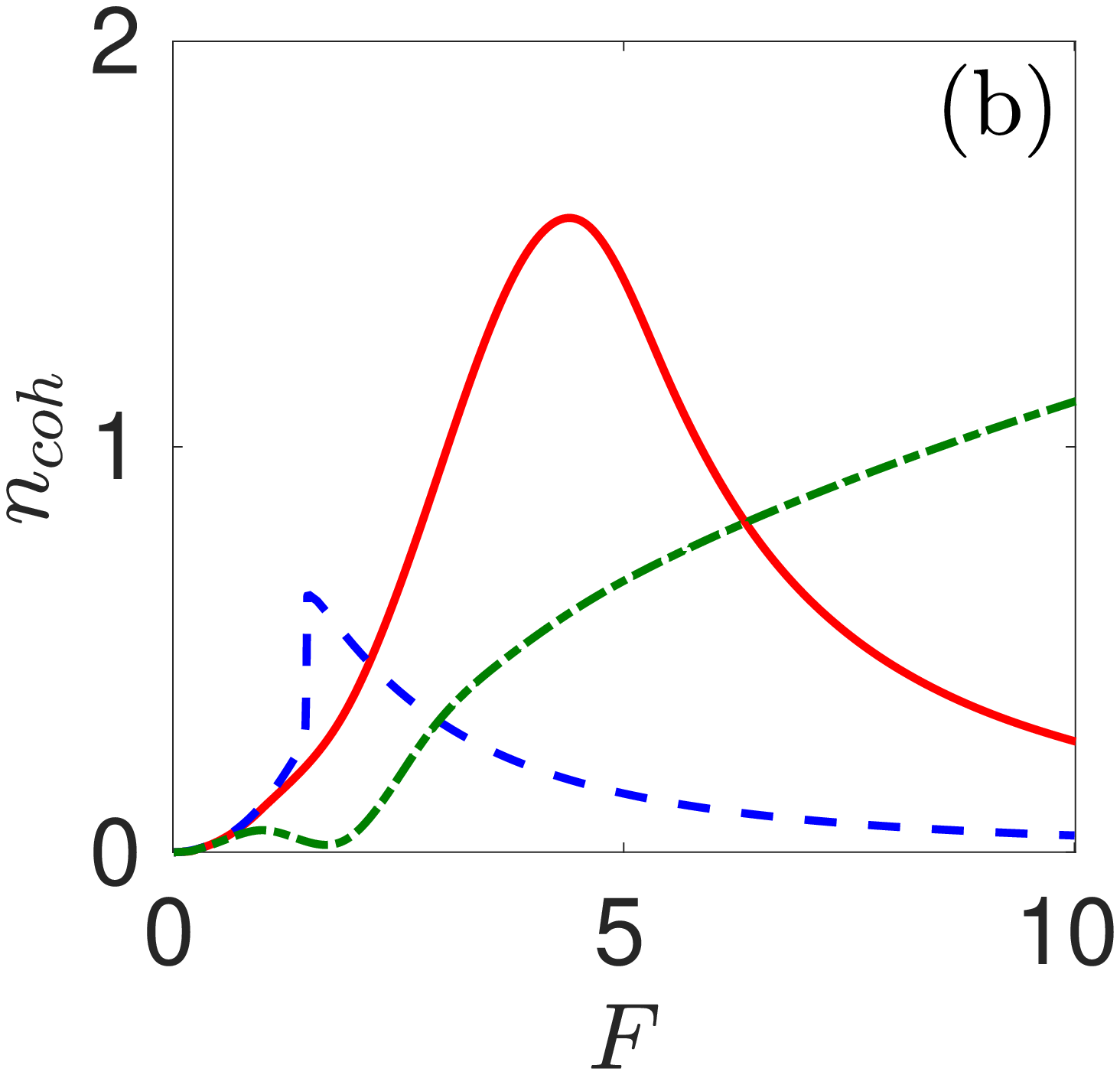} & \hspace{-.2in}
    \includegraphics[scale=0.3]{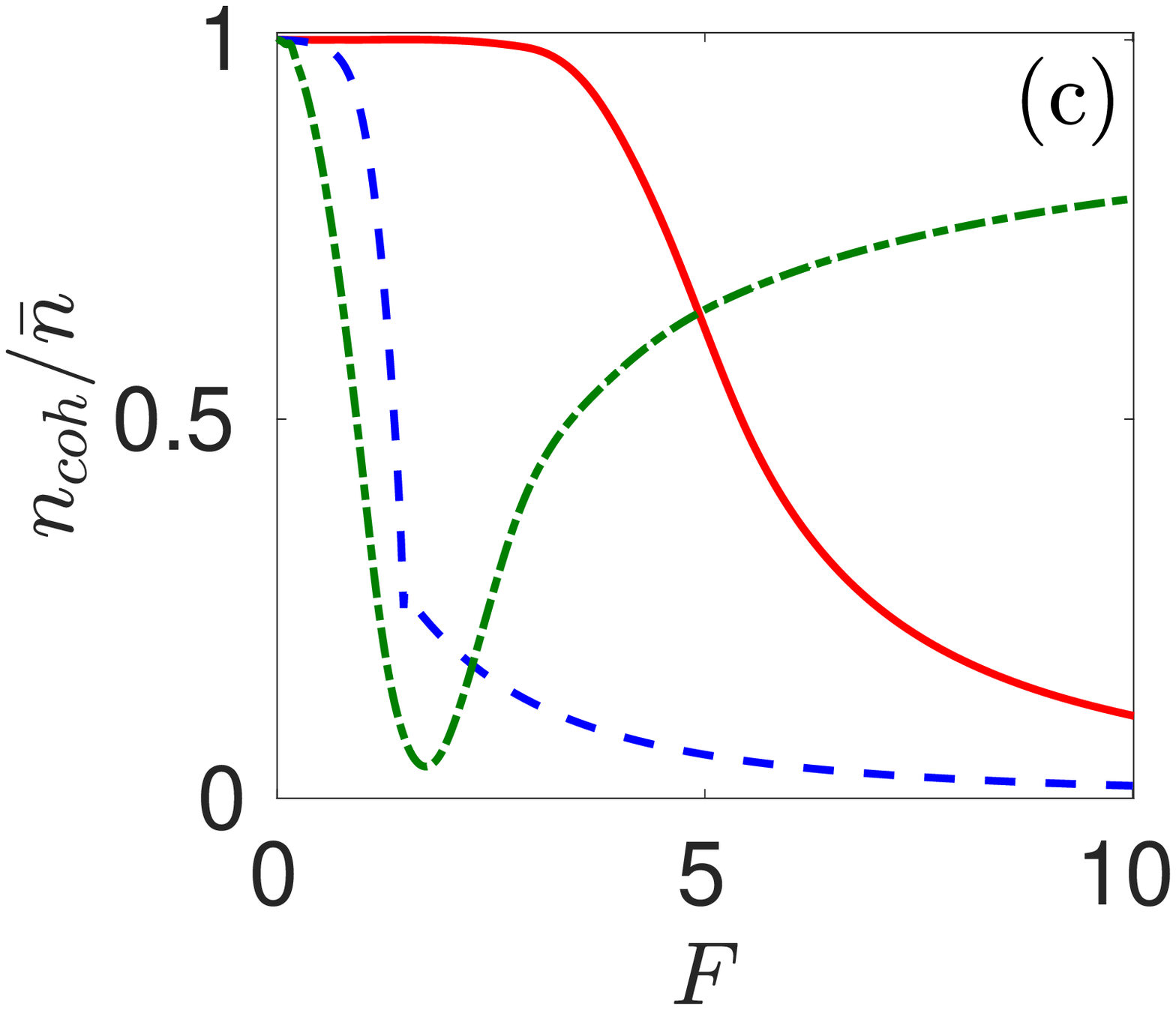}
     \end{tabular} 
 \caption{\small Plotted are the mean particle number $\bar n$ (a), coherent density $n_{\rm coh}$ (b), the coherent fraction  $n_{\rm coh}/\bar{n}$ (c) as a function of $F$ for $\Delta\omega=2$, $J=1$, $U=0.01$ (red solid), $U=1$ (blue dashed) and $U=10$ (green dashed-dotted).}
 \label{nbar}
\end{figure} 
 
  \begin{figure}[h]
	\centering
	\begin{tabular}{@{}ccc@{}}
		\hspace{-.2in}
		\includegraphics[scale=0.3]{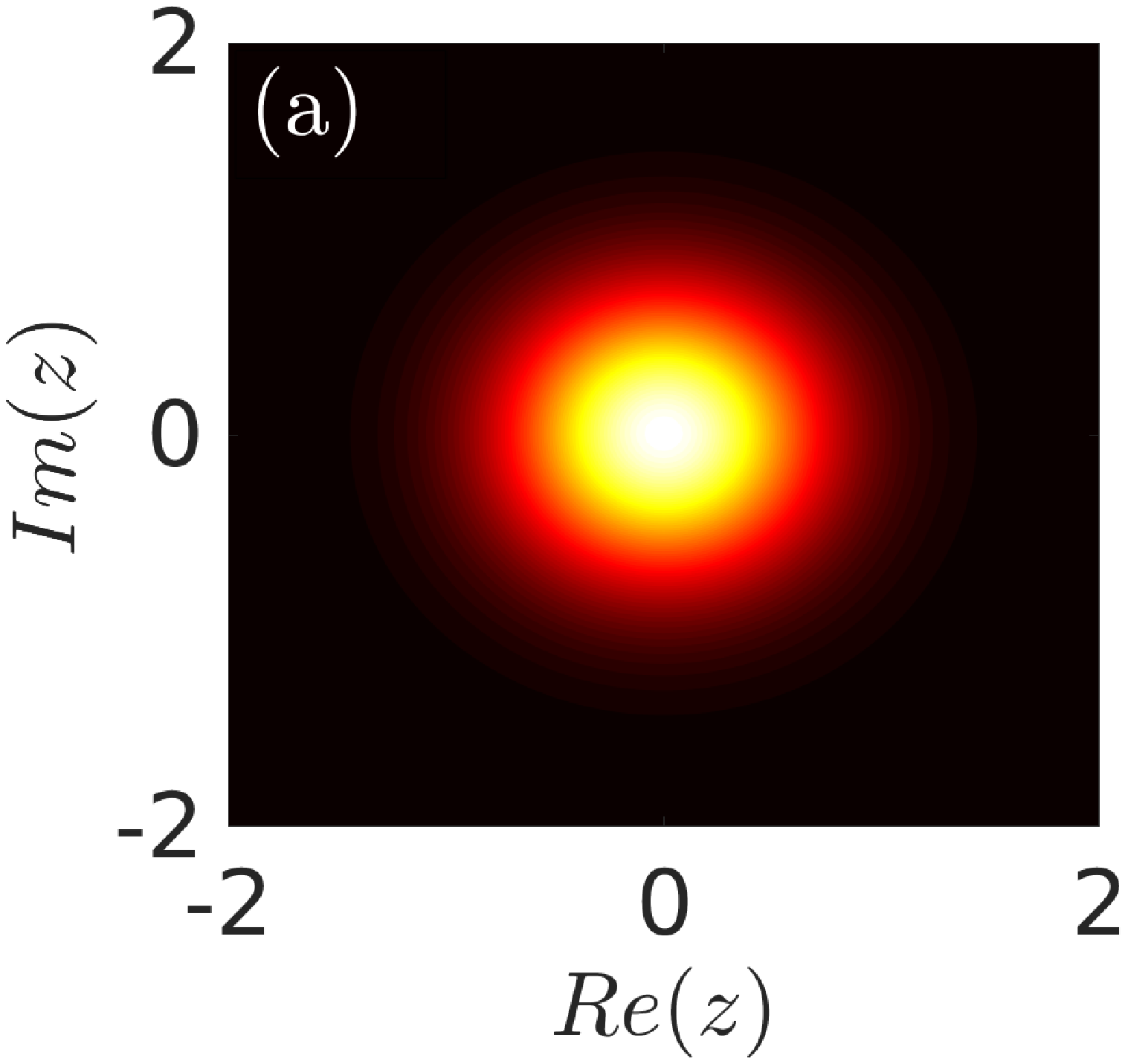} & \hspace{-.2in}
		\includegraphics[scale=0.3]{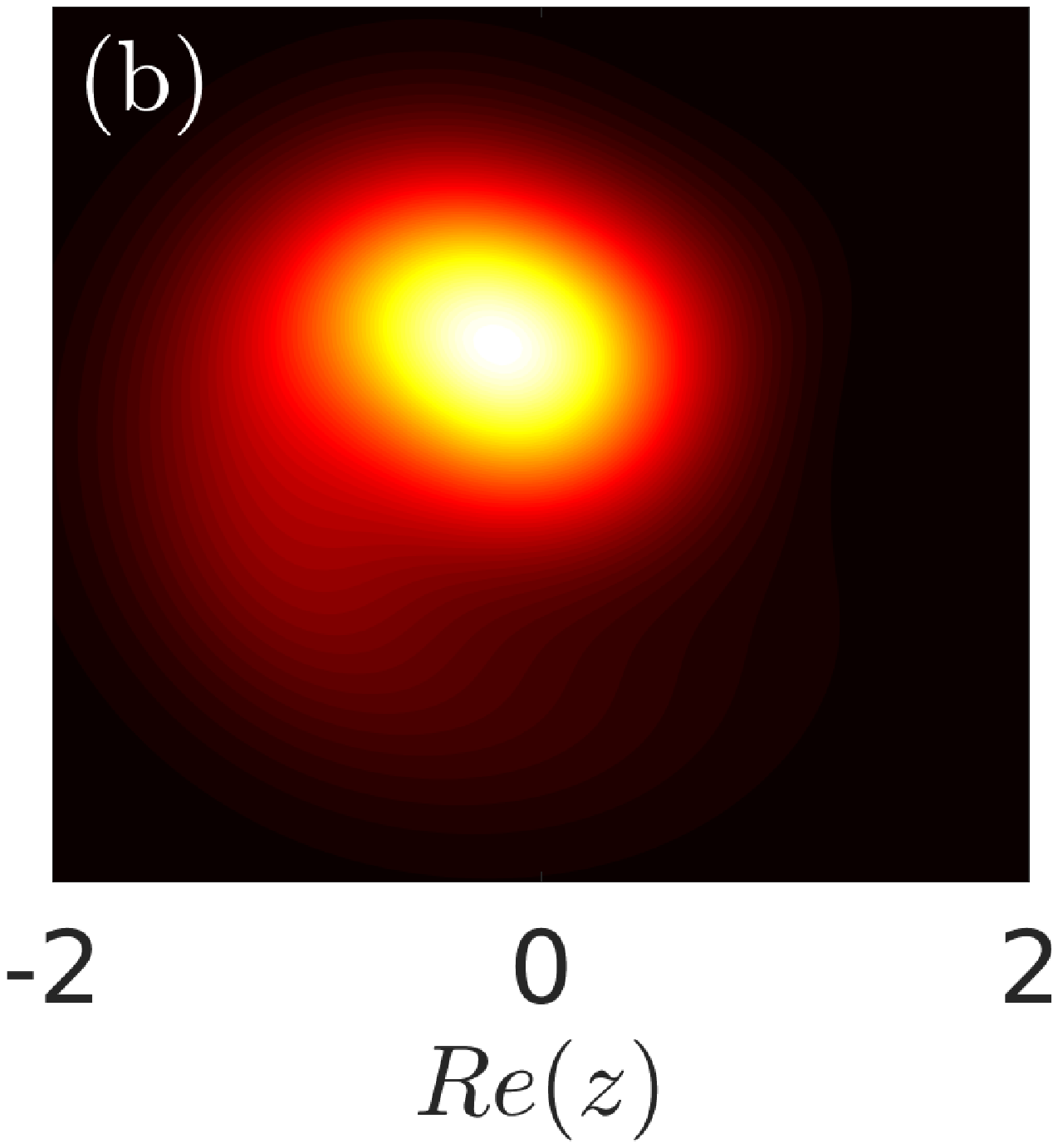} & \hspace{-.2in}
		\includegraphics[scale=0.3]{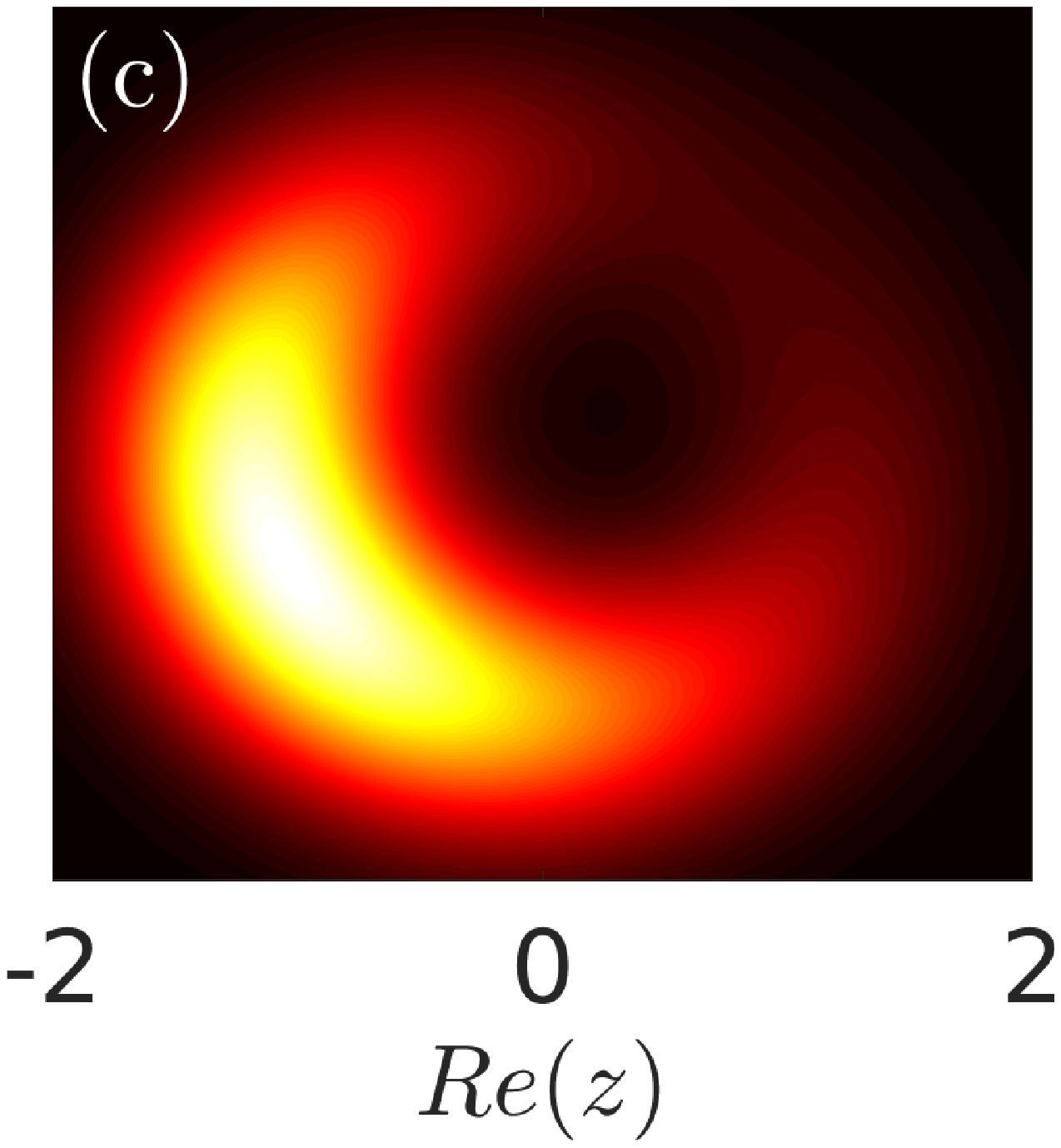}
	\end{tabular} 
	\caption{\small The steady-state Wigner functions $W(z)$ as a function of the real and imaginary parts of the  coherent field $z$ for $J=1$, $\Delta\omega=2$, $U=1$, $F=0.02$ (a), $F=1.17$ (b) and $F=1.8$ (c). The subplots (a) and (c) present Wigner functions below and above the transition point, respectively; while the subplot (b) presents the same at the transition point. White color corresponds to high values, and red corresponds to zero (a different scale is used for the different panels).}
	\label{Wigner}
\end{figure}

\section{Results and discussion}\label{sec3}

For our numerical work, we take $\hbar\Gamma$ as the unit of energy and therefore scale all the energy quantities with this unit.
We first identify the parameter space where bi-stability occurs by analyzing the roots of the semi-classical Eq. (\ref{eq15}). There are three real and positive roots for any set of parameters if it satisfies the condition $\left(J+\Delta\omega\right)>\frac{\sqrt{3}}{2}$, but this  does not happen always. It can be understood by taking the derivative of the left side  of Eq. (\ref{eq15}). We have scanned the solutions of Eq. (\ref{eq15}) for a wide range of system parameters.
Fig. \ref{Figure 4.} shows the mono- and bi-stability ($S$-shaped curve) of the semi-classical mean number density $n_{ mf}$. The system has two kinds of stability: (i) when three roots are real and positive then the system enters into a bi-stable region and (ii) when only one real root is survived then the system becomes mono-stable. In the bi-stable regime, two roots are associated with two high density phases and the remaining root defines low density phase. The semi-classical GP equation (\ref{eq13}) captures only one high-density and one low-density phase which are stable and another high density phase that is always unstable, in consistence with the  generalized {\em P} representation of Drummond and Walls \cite{Drummond}.

 \begin{figure}[h]
	\centering
	\begin{tabular}{@{}ccc@{}}
		\hspace{.2in}
		\includegraphics[scale=0.3]{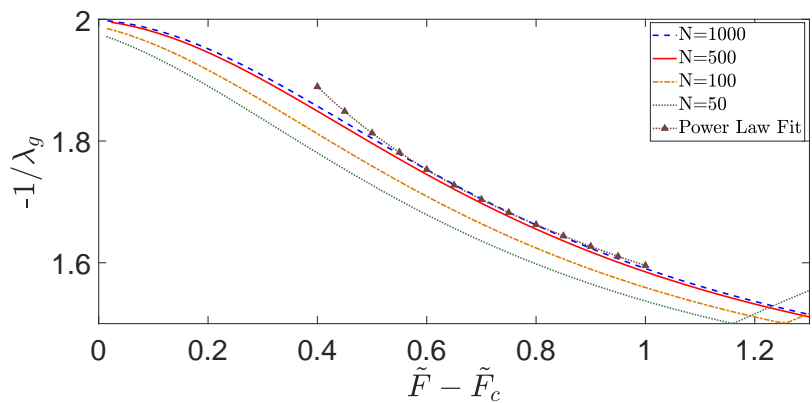} 
	
	\end{tabular} 
	\caption{\small Variation of relaxation time ($-1/\lambda_g$) as a function of $(\tilde F -\tilde F_c)$ for different values of $N$. The parameters are $U=1$ and $\Delta\omega=2$.}
	\label{scaletime}
\end{figure}

\begin{figure}[b]
	\centering
	\begin{tabular}{@{}ccc@{}}
		\hspace{-.2in}
		\includegraphics[scale=0.3]{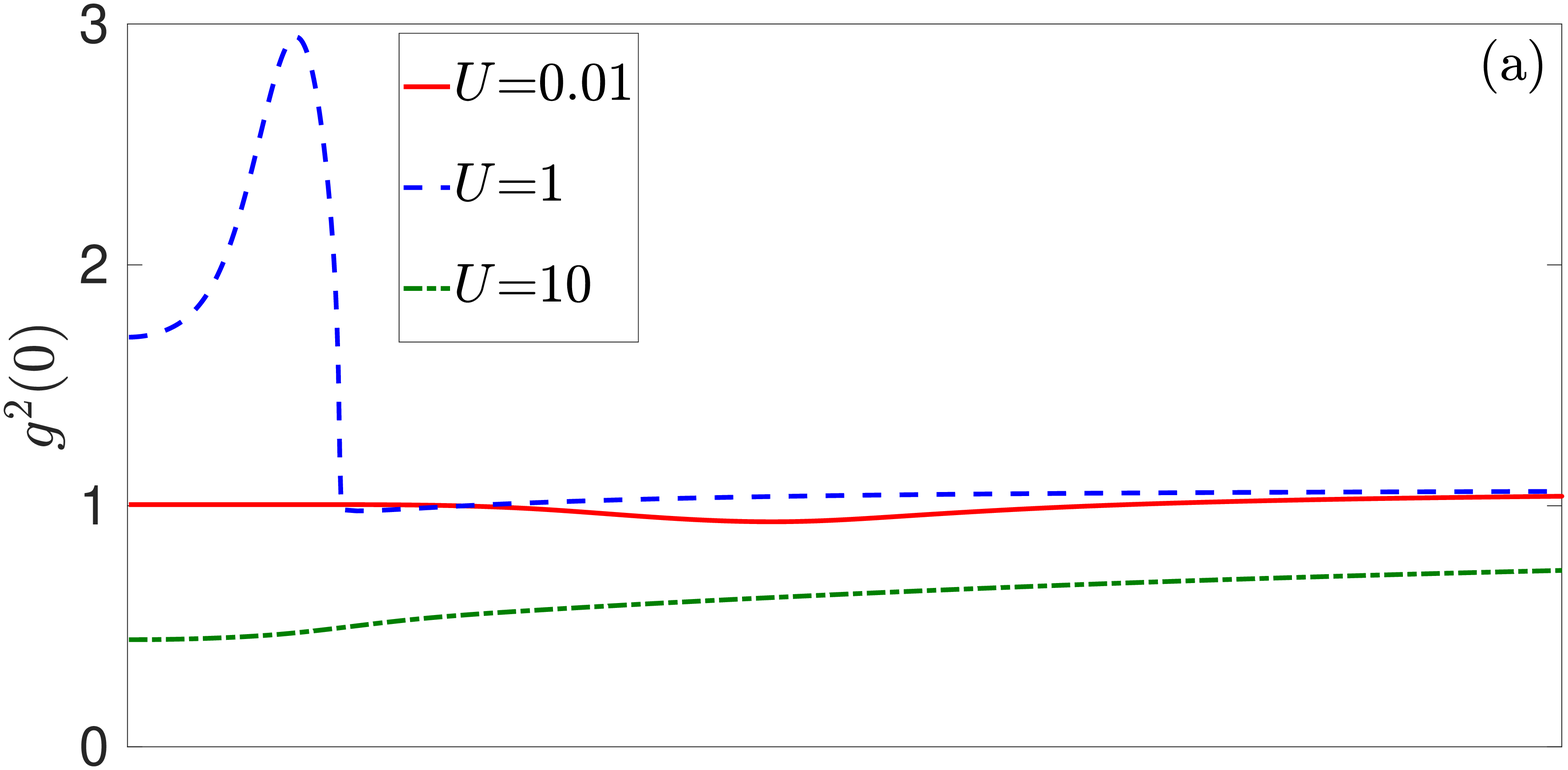}
		\vspace{-0.4in}
	\end{tabular}
	\begin{tabular}{@{}ccc@{}}
		\hspace{-.2in}
		\includegraphics[scale=0.3]{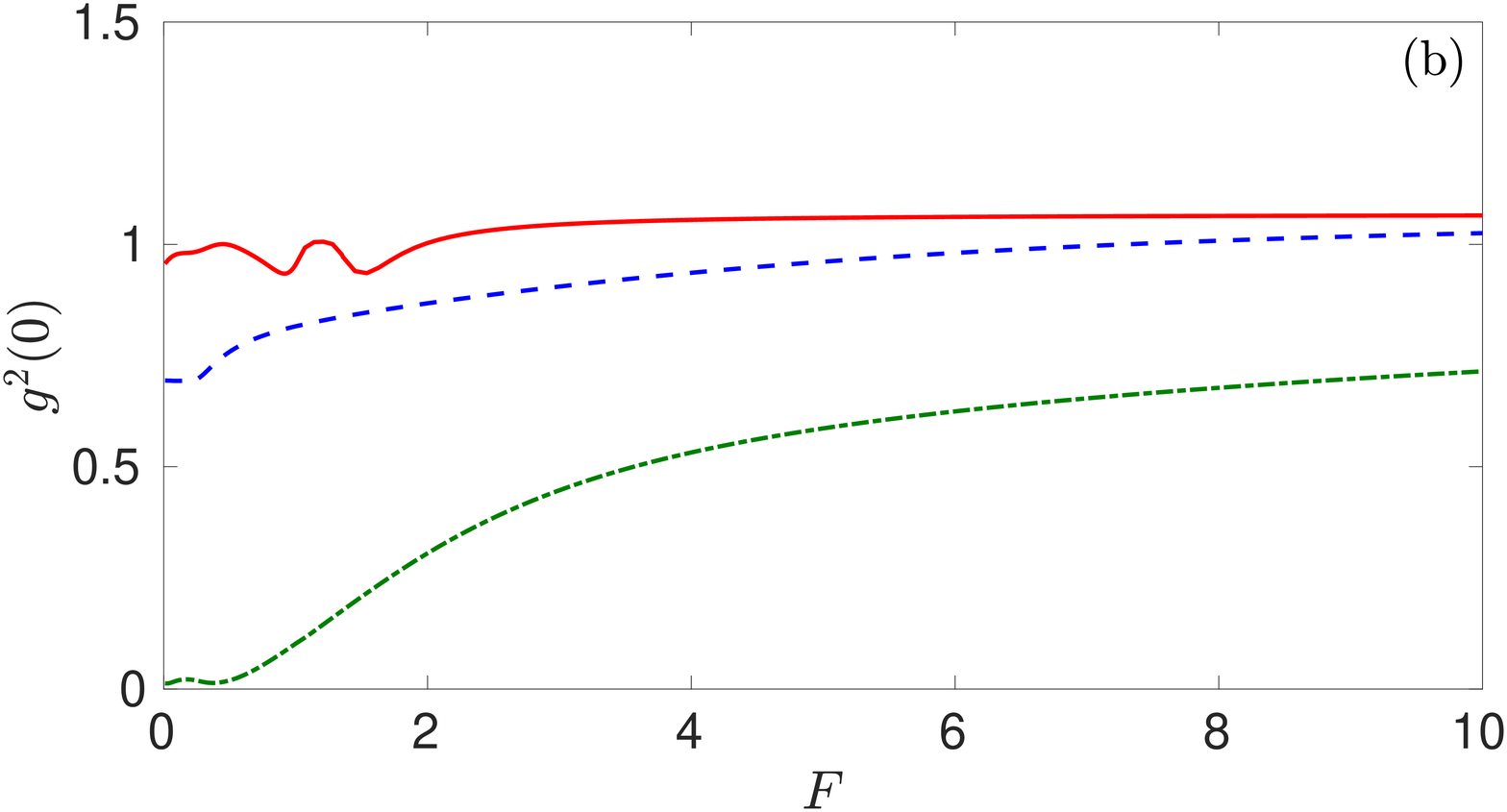} 
	\end{tabular}  
	\caption{\small The normalized equal-time second order correlation function $g^{(2)}(0)$ as a function $F$ for $\Delta\omega=2$ (a) and $\Delta\omega=0.2$ (b). The other fixed parameters are $J=1$,  $U=0.01$ (red solid), $U=1$ (blue dashed) and $U=10$ (green dashed-dotted). }
	\label{g20}
\end{figure}

 Next we carry out full quantum mechanical treatment. We define the mean particle number per lattice site  by $\bar n = \frac{1}{N_{lat}} \sum_j {\rm Tr}\left[\hat b_j^{\dagger}\hat b_j \hat\rho^{ss}\right]=n_{ coh}+n_{ nc}$, where $N_{lat}$ denotes the number of lattice sites, $n_{ coh}$=$|\psi|^2$ is the coherent and $n_{nc}$ is the non-coherent part of the density. At a small on-site repulsion, the system's behavior is dominated by  coherent number fluctuation  in each site but as the ratio $\frac{U}{J}$ increases the on-site number fluctuation drastically reduces. In Fig. \ref{nbar}, we present the variation of the mean particle number $\bar n$ as a function of $F$. In contrast to  the semi-classical treatment depicted in Fig. \ref{Figure 4.}, bi-stable nature is absent in quantum treatment. Instead, we notice that when  $U$ becomes comparable to $\Delta\omega$ there exists a sudden discontinuous jump from one to another semi-classical branch at a critical drive strength $F=F_c$, indicating the onset of a first order DPT. In Fig. \ref{nbar}(a), this critical behaviour occurs at $F_c=1.1$ for $U=1$ and $\Delta\omega=2$.

 We now calculate the steady-state Wigner distribution to further illustrate quantum signature of DPT. It is given by 
\begin{eqnarray}
W(z,z^*)=\frac{1}{\pi^2}\int d^2\nu e^{\nu^* z-\nu z^*} {\rm Tr}\left[\rho^{ss}e^{\nu a^{\dagger}-\nu^*a}\right]
\end{eqnarray}
where $z$ and $\nu$ are coherent states and $\int d^2z W(z)=1$.

A close inspection into steady-state Wigner distribution function displayed in Fig. \ref{Wigner} reveals the signatures of DPT. Below the transition point, the function has a single symmetrical-peak structure that corresponds to the  mono-stable nature  of   the system. This is an evidence for a valid single-valued root of the semi-classical  Eq. (\ref{eq15}). Above the transition point, the well-known bimodal shape appears providing a quantum signature of the  semi-classical bi-stability. Right at the transition point, the shape of the distribution begins to deform indicating  the switch-over from one to the other phase.

 Another way to analyze critical behavior associated with a DPT \cite{Fazio17} is to introduce an equivalence of thermodynamic limit by employing a  dimensionless parameter $N$ and defining the scaled interaction parameter  $\tilde U=NU$ and scaled drive strength $\tilde F=F/\sqrt N$  such that in  the limit  $N\rightarrow\infty$  the quantity  $UF^2$ remains constant; and to study the variation of the inverse of the Liouvillian gap $\lambda_g$ as a function of $\tilde F$ near the critical drive strength $\tilde{F}_c$. In stark contrast to a second order phase transition \cite{sachdev}, a first order DPT has different behavior as $\tilde F$ approaches to $\tilde F_c$ \cite{dpt,Fazio17}. For a first order DPT, it is experimentally found that $-1/ \lambda_g$ as a function of ($\tilde F-\tilde F_c$) shows a power law behavior over a limited range away from $\tilde F_c$ and an exponential decay near $\tilde F_c$ \cite{dpt}. Furthermore, it was experimentally shown that the critical value of the drive strength $F_c$ is that value of $F$ for which the bunching $(g^{(2)}(\tau)> 0)$ has the longest duration \cite{dpt}. Fig. \ref{scaletime} illustrates the scaling behavior of the relaxation time ($-1/ {\lambda_g}$) as a function of $\tilde{F}$ near $\tilde{F_c}$ for the first order DPT corresponding to the dashed curve in Fig. \ref{nbar}(a).
  We have fitted the curve for $N>>1$ with a weighted power law 
\begin{equation}
 -\frac{1}{\lambda_g}\sim \left(\frac{\tilde F -\tilde F_c}{f}\right)^{-\alpha}
\end{equation}
As shown in Fig. \ref{scaletime}, we have found a reasonably good fit with this power law over a limited range of $\tilde F$ slightly away from $\tilde F_c$ with the exponent $\alpha=0.18$ and $f=12.7$. Remarkably, our results show that as $N\rightarrow\infty$, the relaxation time becomes insensitive to $N$.

Next, we present our results on HBT two-particle correlation functions and analyze their characteristic features at and near the DPT point.  In Fig. \ref{g20} we display the steady-state correlation $g^{(2)}(0)$ as a  function of $F$. We notice that when the input parameters $U$, $J$ and $\Delta \omega$ are set at values that correspond to the phase transition point (as per our observations in Figs. \ref{nbar},  \ref{Wigner} and \ref{g20}), $g^{(2)}(0)$ as a function of $F$ shows 
a prominent peak structure (blue-dashed curve of Fig. \ref{g20} (a)). The peak exceeds 2 meaning the occurrence of  strong non-classical fluctuations and super-bunching  at  the transition point. Perhaps, this super-bunching behavior forms  a benchmark of the first order DPT as is also observed earlier by several authors \cite{dpt,Houck,ten2,Fazio17}. For small $U$ (red solid curve),  $g^{(2)}(0)$ as a function of $F$ varies very little near unity implying coherent nature of the two-particle correlation. In contrast, when $U$ is large,  $g^{(2)}(0)$ is much smaller than unity for low values of $F$ meaning that the system has strong anti-bunching character with sub-Poissonian particle distribution.    

 \begin{figure}[t]
\centering
  \begin{tabular}{@{}ccc@{}}
     \hspace{-.2in}
    \includegraphics[scale=0.3]{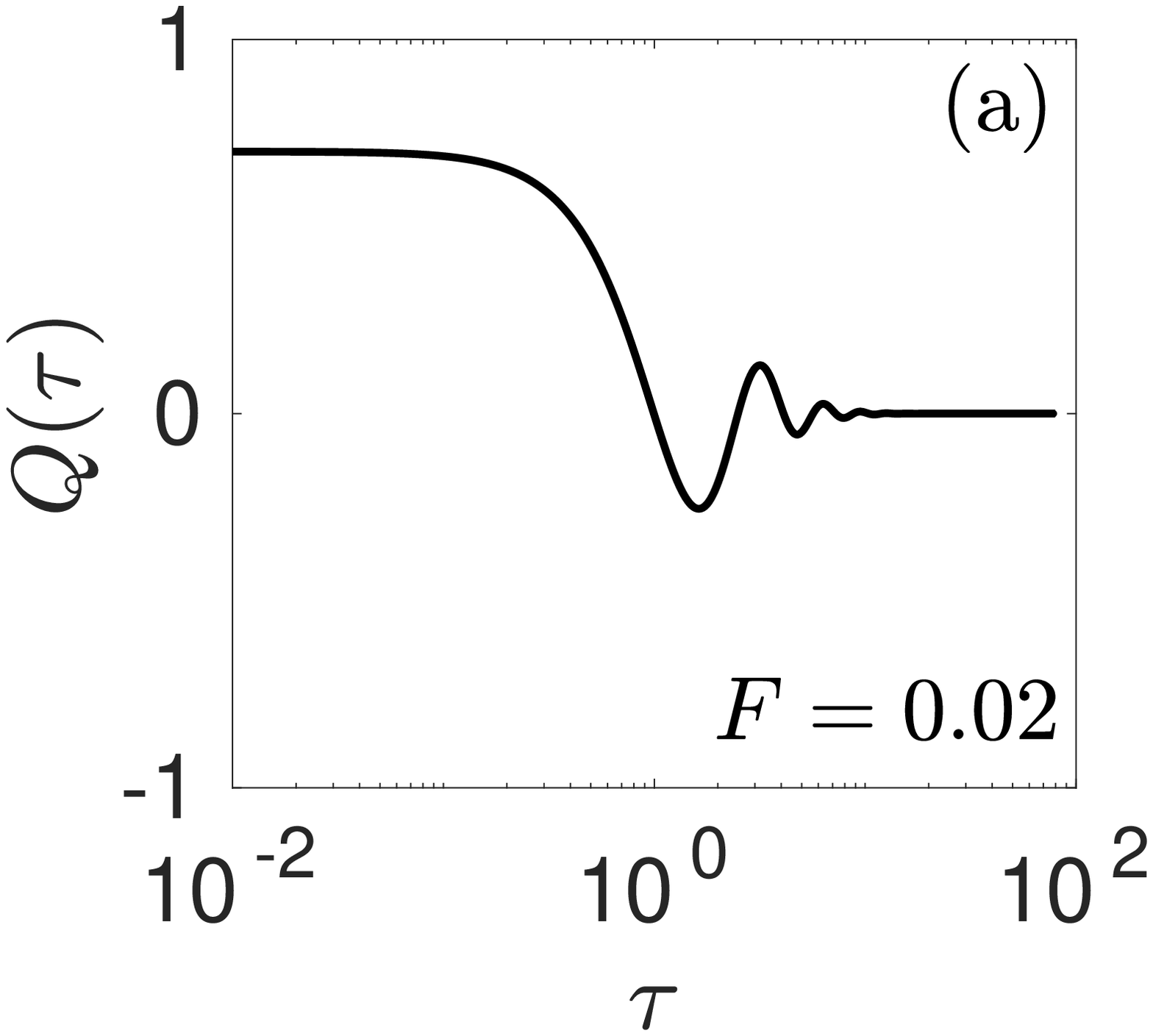} & \hspace{-.15in}
    \includegraphics[scale=0.3]{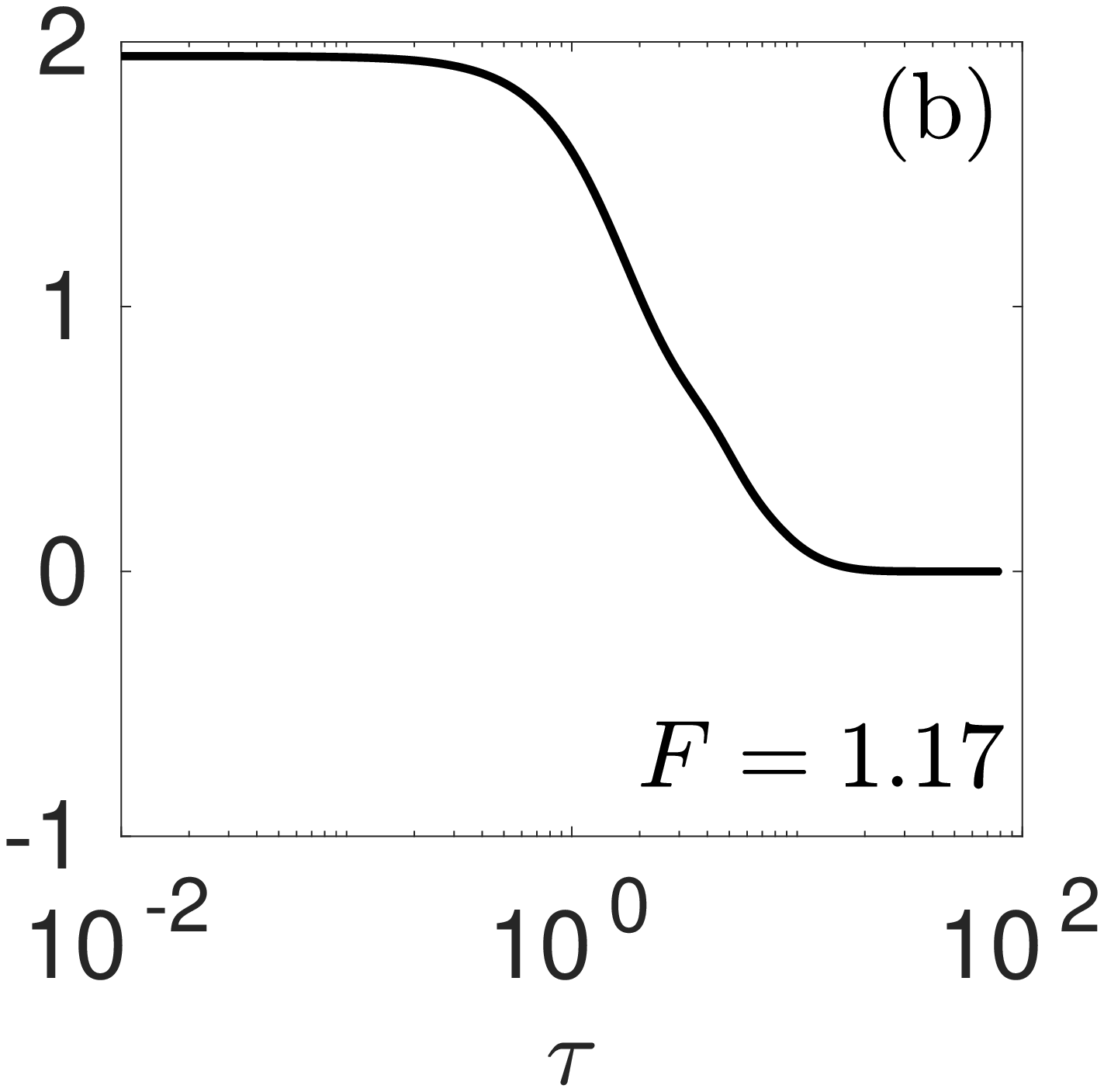} & \hspace{-.15in}
    \includegraphics[scale=0.3]{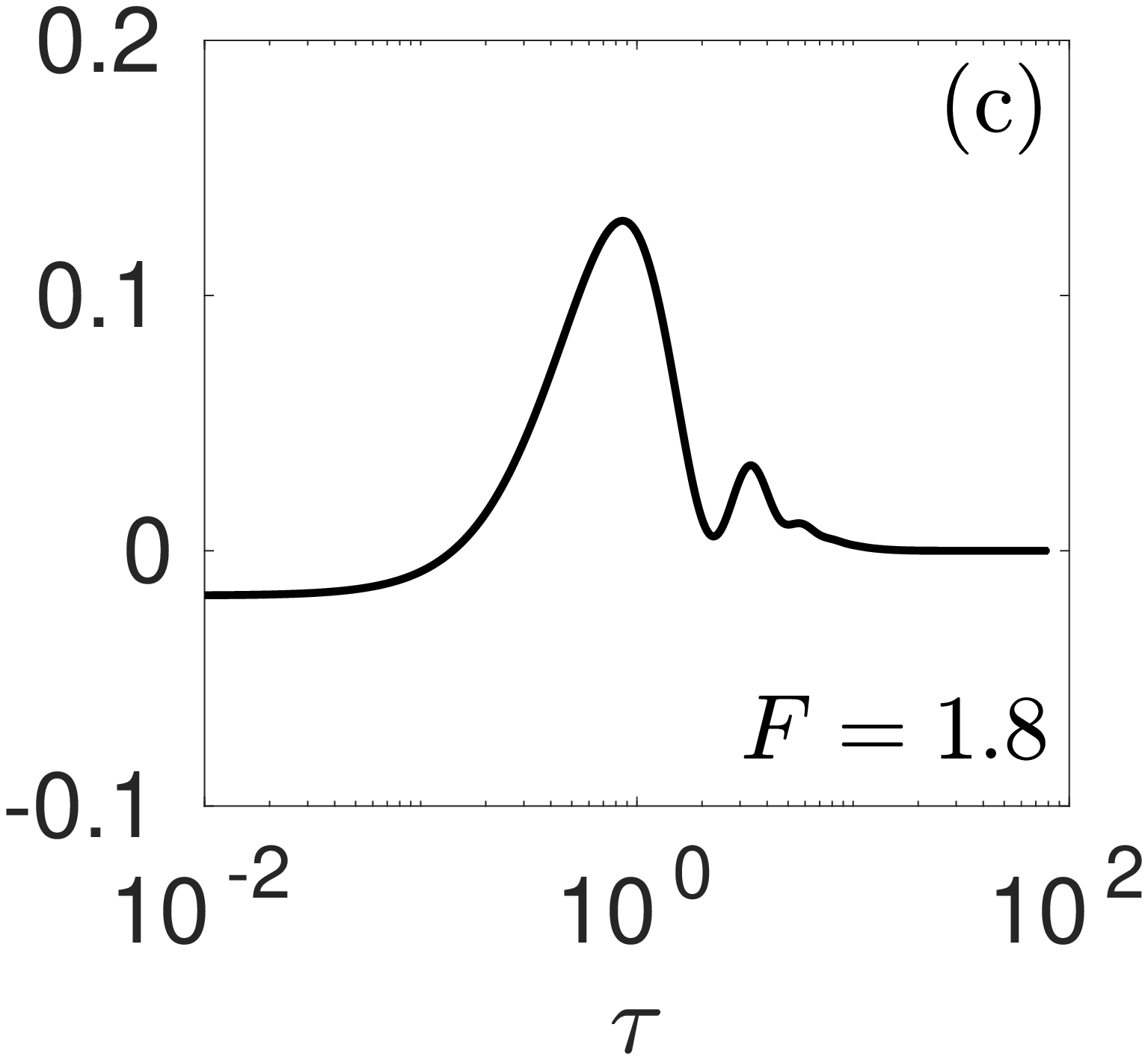} 
    \end{tabular}
    \begin{tabular}{@{}ccc@{}}
    \hspace{-.2in}
    \includegraphics[scale=0.3]{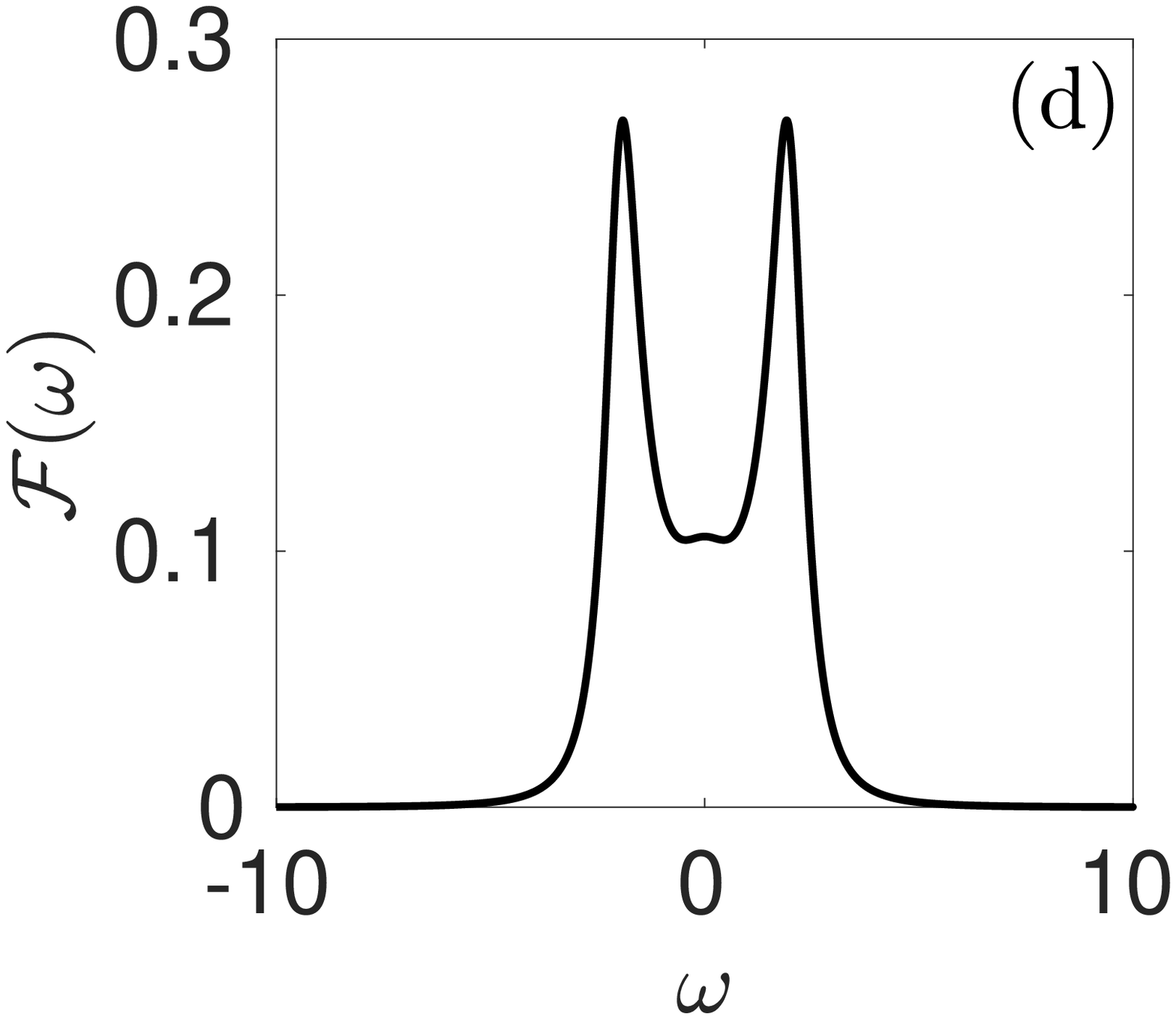} & \hspace{-.15in}
    \includegraphics[scale=0.3]{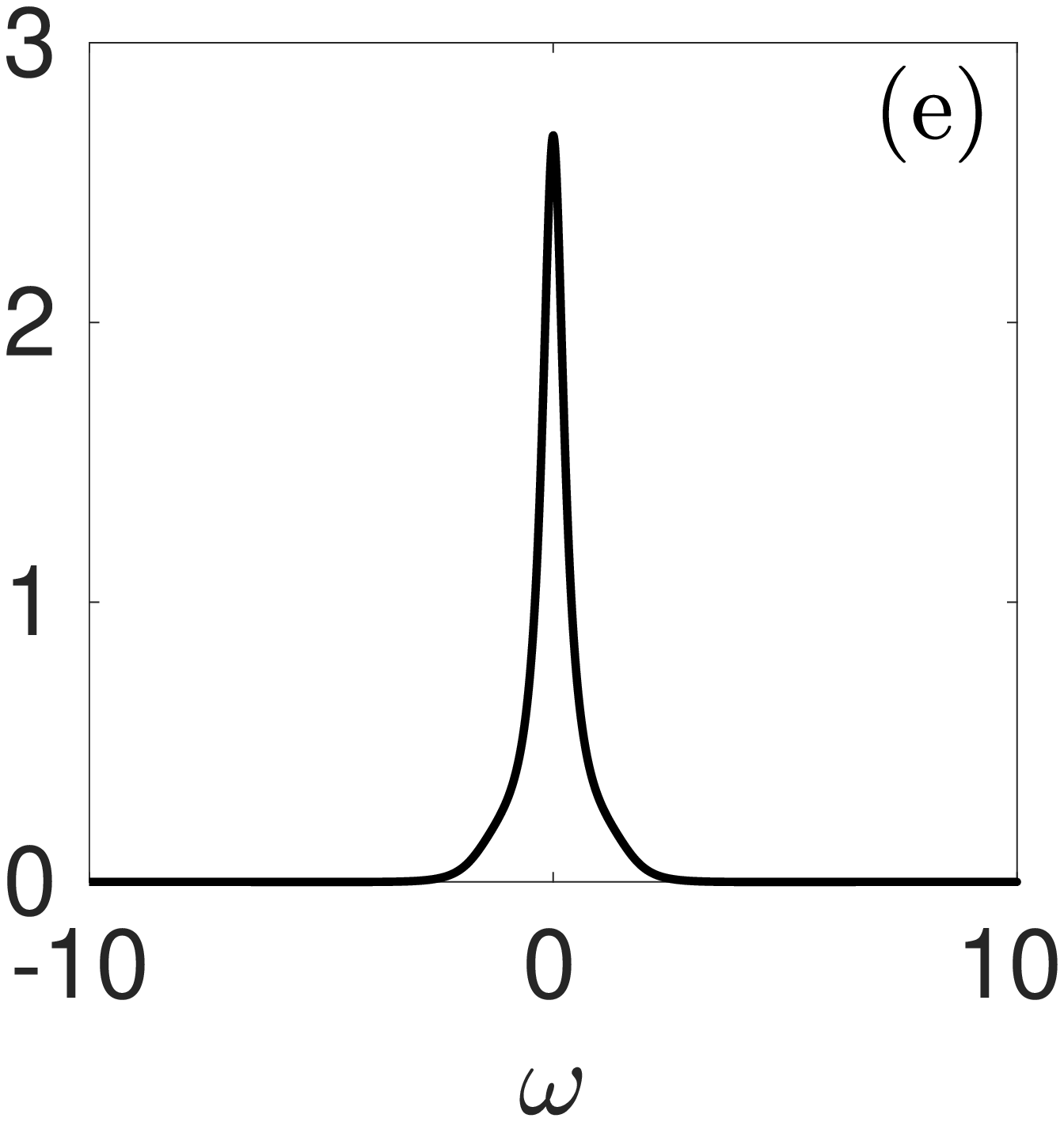} & \hspace{-.15in}
    \includegraphics[scale=0.3]{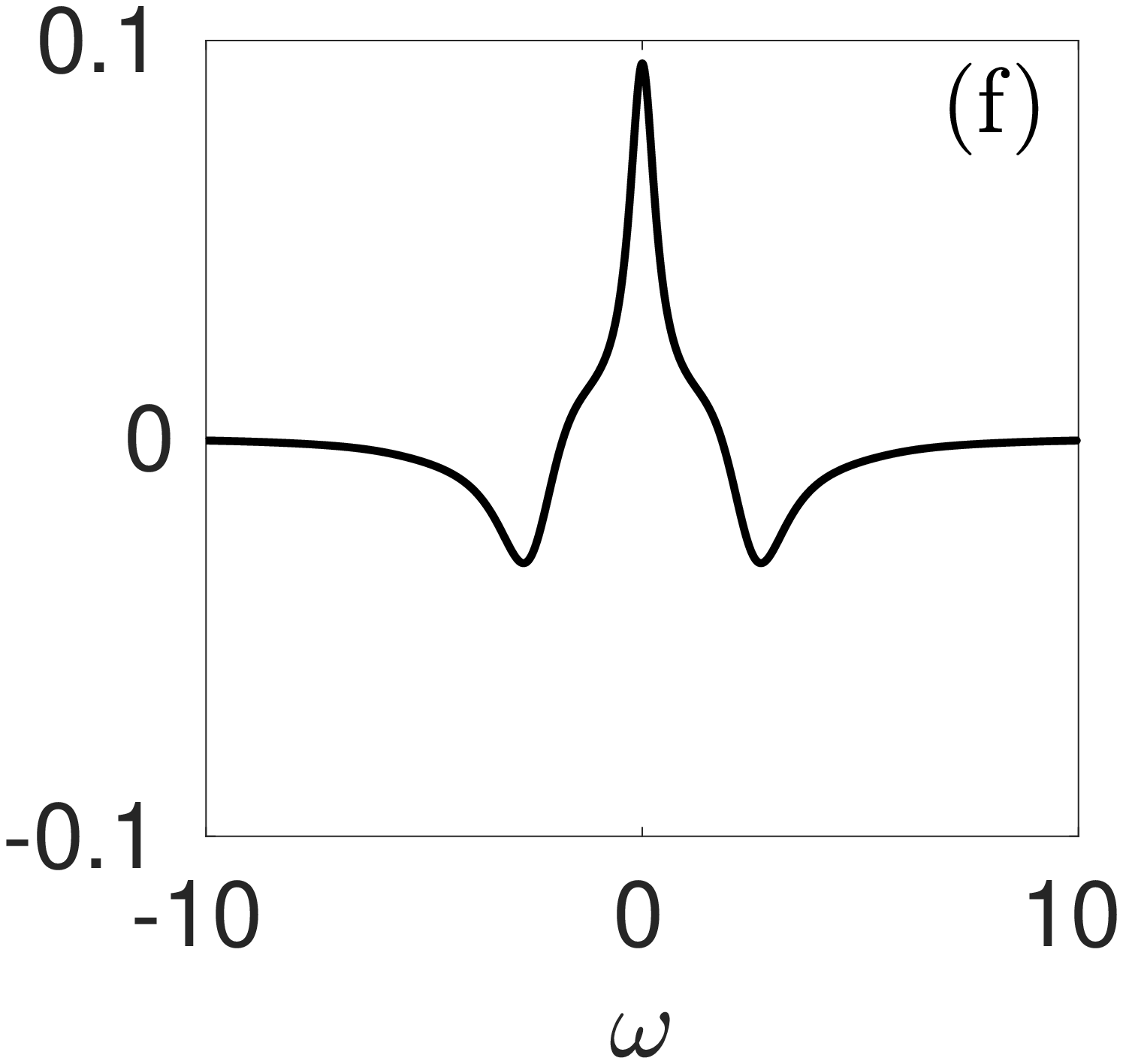}
     \end{tabular} 
 \caption{ The two-time correlation function  $Q(\tau)$ and its  Fourier transform $\mathcal{F}(\omega)$ are plotted as a function of time delay $\tau$ and frequency $\omega$, respectively, for three different values of $F$.  The other parameters are kept fixed at $U=1$, $\Delta\omega=2$ and $J=1$. The Liouvillian eigenvalues associated with the gap is calculated to be   $\lambda_2 = -0.5\pm 1.9998i$ (a, d),  
 	$\lambda_1 = -0.2921 +  0i$ (b, e) 
 and  $\lambda_2 = -0.8575 \pm 3.1229i$ (c, f). The positions of the peaks or dips correspond to the imaginary parts of $\lambda_1$ or $\lambda_2$, while the HWHM of the peaks or dips correspond to the real part of the eigenvalues (see text). }
 \label{g2t}
\end{figure}

\begin{figure}[t]
	\centering
	\includegraphics[scale=0.3]{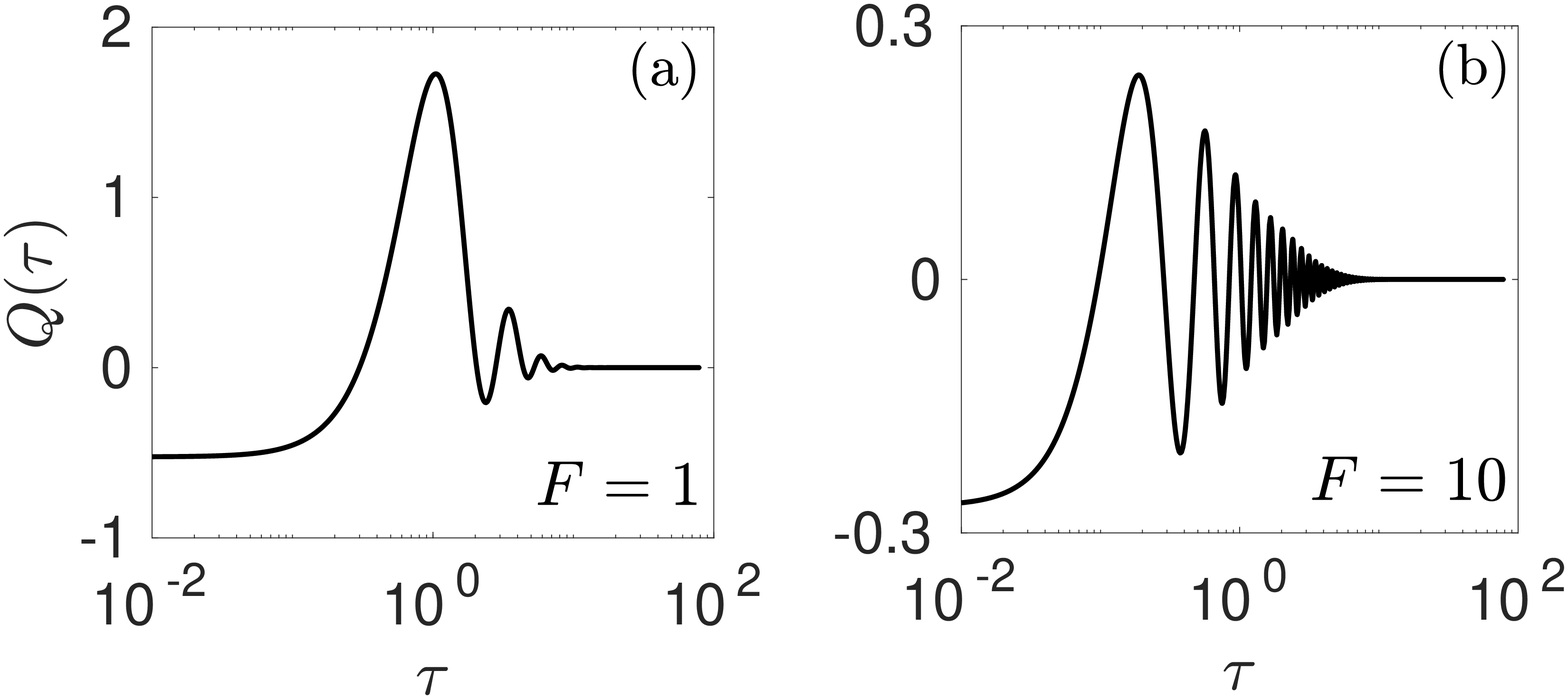} 
	\includegraphics[scale=0.3]{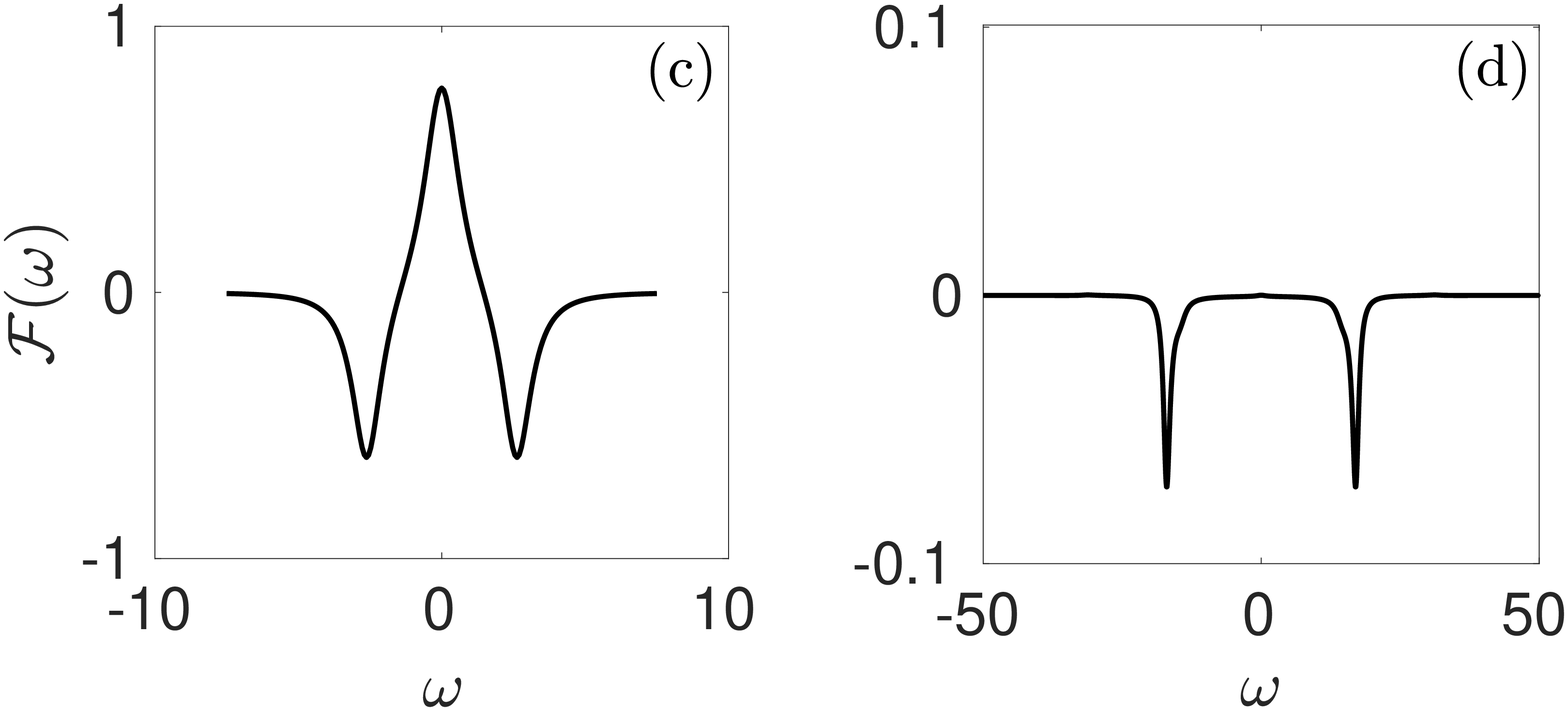} 
	\caption{\small Same as in Fig. \ref{g2t} but for  $U=10$, $\Delta\omega=2$, $J=1$,  $F=1$ (a and c) and  $F=10$  (b and d). For (a) $\lambda_2 =-0.6268 \pm 2.6188 i$ and  for (b) $\lambda_2 =-0.6859 \pm 17.0303 i$. In (c), the two  peaks are located  at $\omega_{\pm} \approx \pm 2.62$. In (d) the  two dips appear at $\omega_{\pm}\approx \pm17$.}
	\label{g2f2}
\end{figure}

The upper panel of Fig. \ref{g2t} shows the temporal  behavior of  $ Q(\tau) = g^{(2)}(\tau) - g^{(2)} (\infty)$ while the lower panel of this figure displays 
its Fourier transform  $\mathcal{F}(\omega)$ as defined in Eq. (\ref{fu15}). 
The subplot \ref{g2t}(b) illustrates the temporal evolution  of two particle correlation $Q(\tau)$ and the subplot \ref{g2t}(e) shows the corresponding frequency-domain correlation  $\mathcal{F}(\omega)$ when the system parameters are set at the transition point. In comparison to other subplots for which the input parameters are chosen away from the transition point, the decay of $g^{(2)}(\tau)$ as shown in Fig. \ref{g2t} (b)   is non-oscillatory and much slower. The corresponding spectral-domain correlation $\mathcal{F}(\omega)$ shown in Fig. \ref{g2t} (e) shows a prominent single-peak structure with the peak lying at zero frequency. We have found that the HWHM of the zero-frequency peak structure is minimum when the parameters are set at the phase transition point. As the subplots (a), (c), (d) and (f) of  Fig. \ref{g2t} illustrate, for parameters away from the phase transition point, $Q(\tau)$ as a function of  $\tau$ exhibits
oscillatory decay and the corresponding frequency-domain correlation $\mathcal{F}(\omega)$ as a function of $\omega$ shows spectral structures that are characteristically quite different from that at the phase transition point. Figure \ref{g2f2} displays again the time- and frequency-domain two-particle  correlations $Q(\tau)$ and $\mathcal{F}(\omega)$ for $U=10$, $\Delta \omega =2$, $J=1$ and two different values of $F=1, 10$. 

\begin{figure}[b]
	\centering
	\begin{tabular}{@{}ccc@{}}
		\hspace{.2in}
		\includegraphics[scale=0.3]{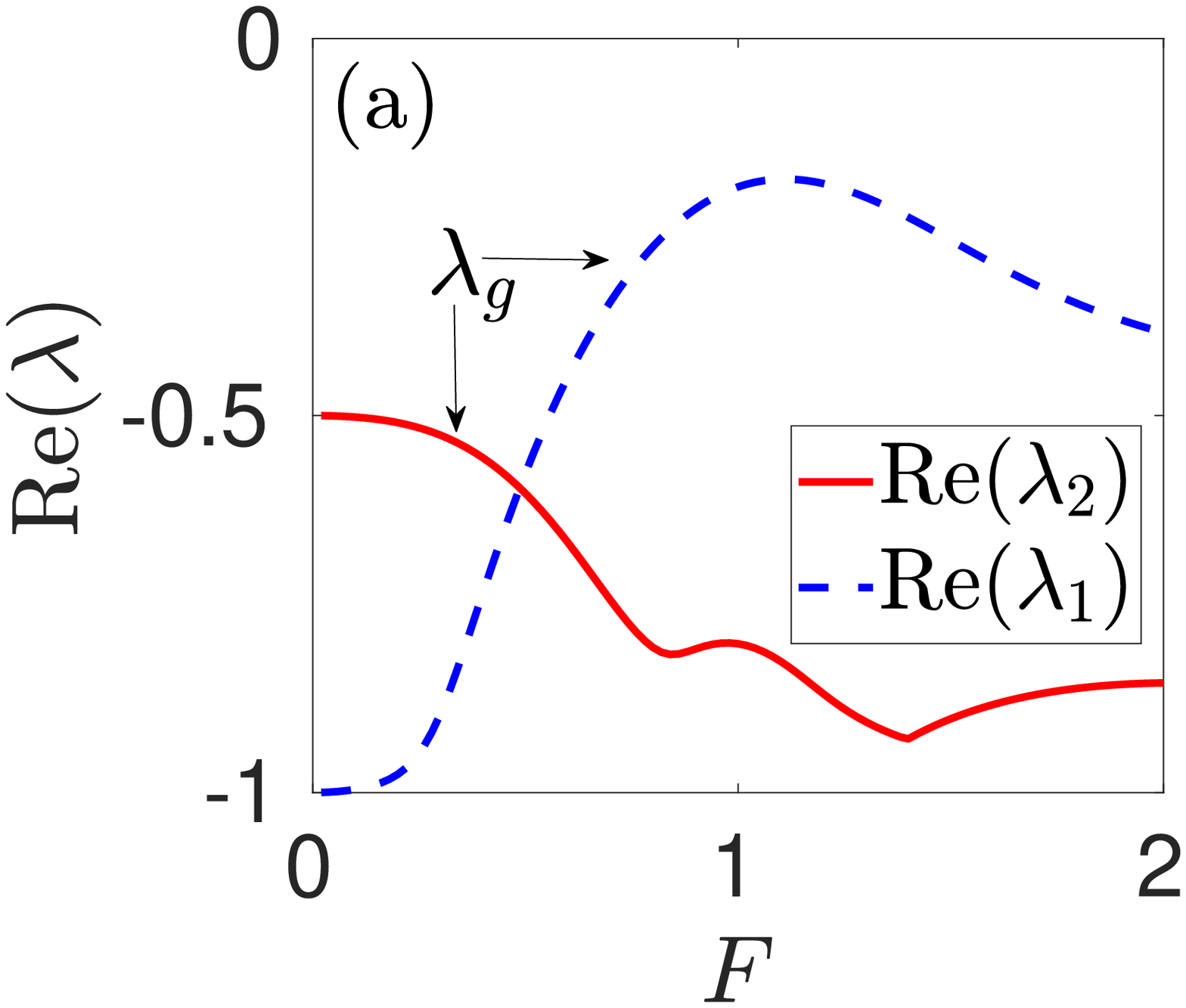} & \hspace{-.1in}
		\includegraphics[scale=0.3]{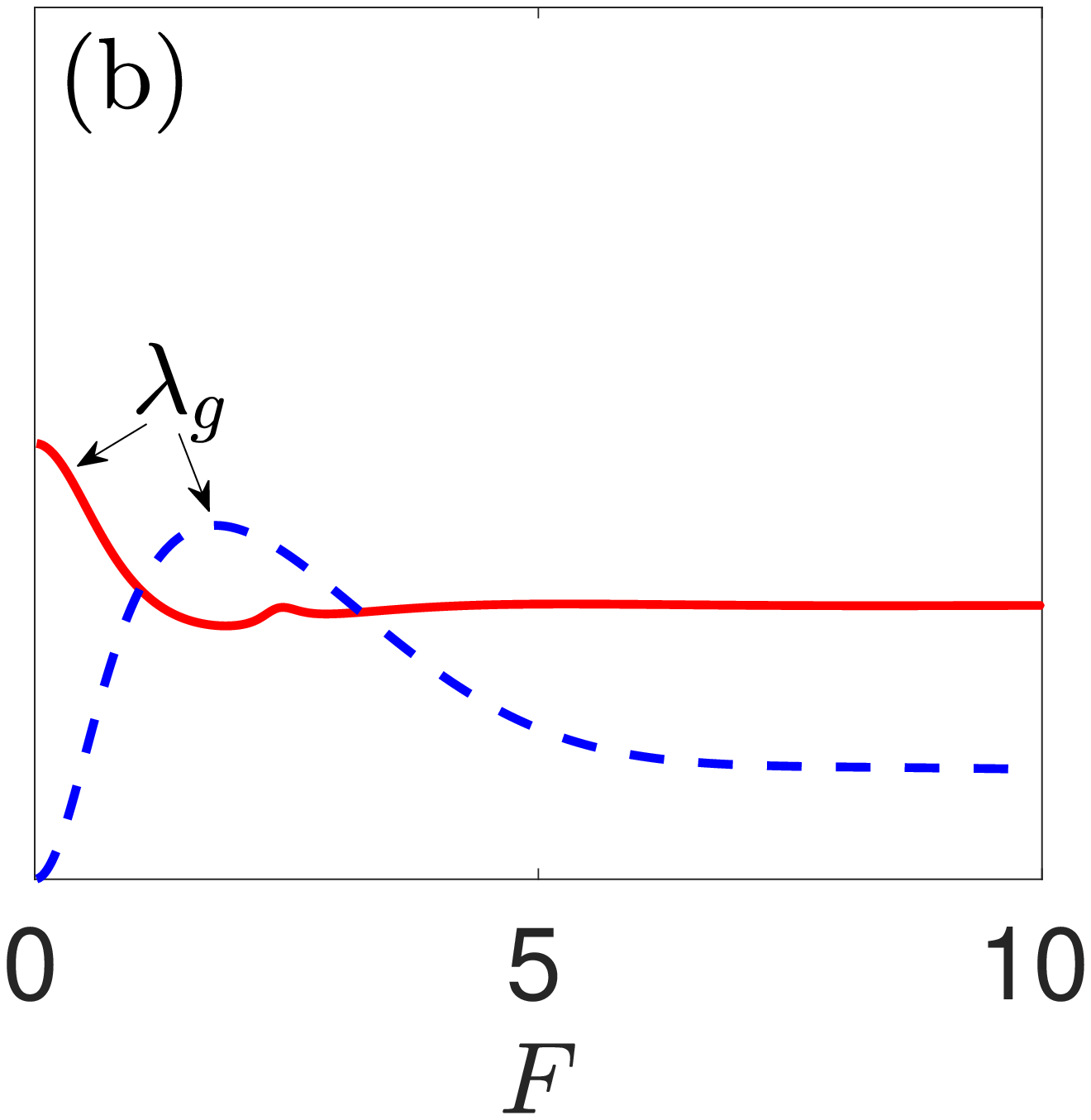}
	\end{tabular} 
	\caption{\small  The real parts of two low lying eigenvalues of the Liouvillian superoperator $\hat{\mathcal M}$ are plotted as a function of $F$ for the parameters (a) $U=1$ and (b) $U=10$. The other fixed parameters are $J=1$ and $\Delta\omega=2$.}
	\label{liou}
\end{figure} 

In contrast to the case of $F=1$, $Q(\tau)$ for  $F=10$ exhibits greater oscillations. The  $\mathcal{F}(\omega)$ for $F=1$ shows a prominent peak at zero frequency and two side dips at frequencies $\pm 2.62$ into the negative regime. In contrast,  $\mathcal{F}(\omega)$ for $F=10$ shows two prominent side dips at frequencies $\pm 17 $ and the peak at zero frequency disappears. The dip structures signify anti-bunching or non-classical nature of the two-particle correlations. Similar dip-like structures have been previously observed in frequency-domain correlations between intensity fluctuations of two optical fields  in the context of  electromagnetically-induced transparency \cite{Martinelli.PhysRevA.94.012503}.

In order to further explain the spectral features observed in Figs. \ref{g2t} and \ref{g2f2}, we have calculated a few low-lying eigenvalues of the Liouvillian super-operator. The frequency-domain two-particle correlation function $\mathcal{F}(\omega)$ is given by Eq. (\ref{fu15}) which is a sum of Lorenzian functions with different spectral weight factors. The eigenvalue which has minimum non-zero real part (in absolute magnitude) is denoted 
as $\lambda_g$ or Liovillian gap. In Fig. \ref{liou} we have plotted the real part of two eigenvalues as  continuous functions of $F$. The blue dashed curve corresponds to the eigenvalue with zero imaginary part while the red solid curve corresponds to non-zero imaginary parts. Since complex eigenvalues appear in complex conjugates, the red curve corresponds to two equal and opposite imaginary parts.  
Let us denote the eigenvalues with non-zero imaginary part but with finite real part as $\lambda_{2}$ while the eigenvalue with zero imaginary part as $\lambda_1$. Figure  \ref{liou}(a)  shows that the ${\rm Re}[\lambda_1]$ and
${\rm Re}[\lambda_2]$ as a function of $F$ have a crossing point at a low value of $F$. So, below the crossing point, the red solid curve is the Liouvillian gap, but above the crossing point, the blue-dashed curve serves as the gap. So, below the crossing point, the eigenvalue corresponding to the red-dashed curve having equal and opposite nonzero imaginary parts primarily determines the nature of the spectral features in $\mathcal{F}(\omega)$ vs. $\omega$ curves. Above the crossing point but $F$ and $U$ being not very large, the spectral features are primarily determined by the Liouvillain gap with zero imaginary part. For large $U$, as  Fig.  \ref{liou}(b) shows,  ${\rm Re}[\lambda_1]$ and
${\rm Re}[\lambda_2]$ have two crossing points. So, for $F$ ranging between the two crossing points, both zero- and nonzero frequencies will dominate in the spectrum as the Fig. \ref{g2f2} (c) indicates while for large $F$, zero-frequency part will be suppressed as Fig. \ref{g2f2} (d)
illustrates. The positions of the spectral peaks or dips are found to coincide  with the imaginary parts of $\lambda_1$ or $\lambda_2$  while the HWHM of the peak or dip structures is given by the real part of $\lambda_1$ or $\lambda_2$.

\section{conclusions} \label{sec6}

In conclusion, we have studied  the time- and frequency-domain HBT  two-particle correlations of a driven dissipative Bose-Hubbard model (BHM) 
and analyzed in detail the various temporal and spectral features of the correlation that reflect quantum statistical properties of the system at, below and above the DPT of the model. Our results show that except at or very near to the phase transition point, $g^{(2)}(\tau)$ in general exhibits oscillatory decay leading to multiple peak- or dip-structures in the correlation function in the frequency domain. The details of spectral structures such as the central frequencies of the peak- or dip-structures and their widths are explained in terms of the Liouvillian eigenvalues and eigenfunctions. We have shown that right at the phase transition point, the correlation spectrum has a single Lorenzian  with zero central frequency and minimum HWHM. Our results further show that the quantum statistical properties of the steady-state can be controlled by tuning on-site interaction $U$, the detuning $\Delta \omega$ and the drive strength $F$. For small $U$ and small $F$ the system at steady-state exhibits coherent or bunching behavior while strongly driven steady-state in the strong interaction regime ($U >\!> 1$) can exhibit strong anti-bunching or strongly correlated phase.  In this paper, we have carried out our investigation under a homogeneous mean-field approximation. Going beyond this approximation and taking into account spatial inhomogeneity in a driven dissipative many-body system will be important step forward to explore an interplay between  HBT and density-density or current-current correlations of the model, which we hope to address in our future communications.

\begin{acknowledgments}
One of us (SM)  acknowledges a support from  Council of Scientific and Industrial Research, Govt. of India. 
\end{acknowledgments}

\appendix
\section{solutions of the steady state density matrix} \label{a1}
In this appendix we present the method of numerical solutions of density matrix elements. Under the decoupling approximation, the density matrix $\hat{\rho} = \prod_j \rho(j)$,
where $\rho(j)$ is the density matrix for the $j$-th site. But under the same
approximation, $\rho(j)$ is site-independent and so we solve for a single-site
density matrix $\rho^{ss}$ in steady-state.
It then follows from Eq. (\ref{eq8}) that the elements $\rho^{ss}_{m,n}$ in steady-state can be cast into a set of linear coupled algebraic equations which can be expressed in the matrix form
\begin{equation}
 \left[X\right]_{N^2\times 1}=\left([A]_{N^2\times N^2}\right)^{-1}[B]_{N^2\times 1} 
\end{equation}
where $A$ is a square matrix containing  coefficients of $\rho_{m,n}^{ss}$,  $X$ is a row matrix that has components  $\rho_{m,n}^{ss}$ and $B$ is another matrix which hold the steady state information.

We can write the steady state density matrix $\hat\rho^{ss}$ in following way
\begin{equation}
 \hat\rho^{ss}=\sum_{m,n}\rho_{m,n}^{ss}|m\rangle\langle n|
\end{equation}

The exact value of  $\psi$ is evaluated by numerically in a self-consistent manner resulting in   
\begin{eqnarray}
 \psi&=&\langle\hat b\rangle\nonumber={\rm Tr}\left[\hat b \hat\rho^{ss}\right] = \sum_n\sqrt{n+1}\rho_{n+1,n}^{ss}\\\nonumber
\end{eqnarray}

The steady state density matrix elements are calculated from the set of coupled equations

\begin{eqnarray}
&-&i\left[\beta\left(\sqrt{m}\rho_{m-1,n}^{ss}-\sqrt{n+1}\rho_{m,n+1}^{ss}\right)+\beta^*\left(\sqrt{m+1}\rho_{m+1,n}^{ss}-\sqrt{n}\rho_{m,n-1}^{ss}\right)\right]\nonumber\\
&-&i(m-n)\left[\frac{U}{2}(m+n-1)-\Delta\omega\right]\rho_{m,n}^{ss}\nonumber\\&+&\frac{\Gamma}{2}\left[2\sqrt{(m+1)(n+1)}\rho_{m+1,n+1}^{ss}-(m+n)\rho_{m,n}^{ss}\right]=0
\label{eq10}
\end{eqnarray}

\section{Frequency-domain HBT correlation} \label{a2}

Using eigenvalue decomposition of $\hat{\mathcal M}$, we obtain 
\begin{equation}
 \int_{-\infty}^{\infty} g^{(2)}(\tau) \exp[ i \omega \tau] d\tau = -  \sum_{\mu = 0}^{N^2} W_{\mu} 
 \left [ \frac{1 }{i(\omega + \lambda_{\mu i}) - |\lambda_{\mu r}|}  + \frac{1}{i(- \omega + \lambda_{\mu i}) - |\lambda_{\mu r}|}\right ]
 \label{fgt}
\end{equation}
where $\lambda_{\mu r}$ and $\lambda_{\mu i}$ stand for the real and imaginary part, respectively, of the eigenvalue $\lambda_{\mu}$. The weight factor $W_{\mu}$ can be calculated in the following way. The $N^2 \times N^2$ matrix ${\mathbf D}$ that diagonalizes the matrix $\hat{\mathcal M}$ can be constructed as an array of the column vectors $u^{\mu}$ in the form ${\mathbf D} = \left [ u^{0} \hspace{0.2cm} u^{1}, \hspace{0.2cm} u^{2} \cdots 
u^{N^2} \right ]$. So, we can write 
\begin{equation}
e^{\hat{\mathcal M} \tau} = {\mathbf D} \cdot {\rm diag} \left [ \exp(\lambda^0 \tau) \hspace{0.2cm} \exp(\lambda^1 \tau) \hspace{0.2cm} \exp(\lambda^2 \tau) \cdots \exp(\lambda^{N^2} \tau)  \right ] \cdot {\mathbf D}^{-1}
\end{equation}
where ${\rm diag}\left [ \cdots \right ]$ stands for diagonal matrix. Let ${\mathbf X}(\tau) = e^{\hat{\mathcal M} \tau}$. Then we can express ${\mathbf X} =  \sum_{\nu, \nu'} X_{\nu \nu'} \mid \nu \rangle \langle \nu' \mid $ where the element $X_{\nu \nu'}$ can be expressed as 
$X_{\nu \nu'} = \sum_{\mu} {\mathbf D}_{\nu \mu} \exp[\lambda^{\mu}\tau] {\mathbf D}_{\nu' \mu}$.  Here $\nu$ and $\nu'$ run from 1 to $N^2$. Since the vectors $u^{\mu}$ can be expressed in Fock basis : $u^{\mu} \equiv u^{\mu}_{n m} \mid n \rangle \langle m \mid$, where $n, m = 1, 2, \cdots N$, the matrix ${\mathbf X}$ can also be written in terms of Fock basis operators by making a proper correspondence between the operators 
$\mid \nu \rangle \langle \nu' \mid $  and  the Fock-basis operators  $\mid n \rangle \langle m \mid$. 
Thus one can calculate the weight factor 
\begin{equation}
W_{\mu} = \frac{{\rm Tr}\left[\hat b^{\dagger}(0)\hat b(0) \sum_{\nu, \nu'} {\mathbf D}_{\nu \mu}  {\mathbf D}_{\nu' \mu} \mid \nu \rangle \langle \nu' \mid  \left(\hat b(0)\hat u^0 \hat b^{\dagger}(0)\right)\right]}{\left({\rm Tr}\left[\hat b^{\dagger}(0)\hat b(0) u^0 \right]\right)^2}
\end{equation}

In the limit $\tau \rightarrow \infty$ only nonzero  contribution to $g^{(2)}(\infty)$ will come from the term associated with zero eigenvalue, 
that is,  $\lambda_{\mu}$ with $\mu = 0$. $g^{(2)}(\infty)$ then reduces to $W_0$.   
Setting $\lambda_{0i} =0$ and taking the limit $|\lambda_{0r}| \rightarrow 0_+$, and using the relation 
\begin{equation}
{\rm lim}_{\epsilon \rightarrow 0} \frac{ 1}{x \pm i \epsilon} = {\mathcal P}(x) \mp i \pi \delta(x)  
\end{equation}
where ${\mathcal P}$ stands for principal value, we obtain 
\begin{eqnarray}
{\rm lim}_{|\lambda_{0r}| \rightarrow 0} {\rm Re}\left [ \frac{W_0}{i\omega  - |\lambda_{0r}|} \right ] = - \pi W_0 \delta(\omega)
\label{fto}
\end{eqnarray}
Substituting Eqs. (\ref{fgt}), (\ref{fto}) in Eq. (\ref{ft}), using the fact that except for $\mu = 0$ all the eigenvalues appear 
in complex conjugate pairs, we obtain
\begin{equation}
 {\cal F}(\omega) =   2 \Gamma \sum_{\mu = 1}^{N^2} \left [ \frac{W_{\mu} |\lambda_{\mu r}| }{(\omega + \lambda_{\mu i})^2 + \lambda_{\mu r}^2 } \right ]
\end{equation}


\bibliographystyle{apsrev4-2.bst}
\bibliography{revised}

\end{document}